\documentclass[a4paper,10pt,twoside]{cpc-hepnp}
\usepackage{multicol}
\usepackage{graphicx}
\usepackage{booktabs}
\usepackage{amssymb,bm,mathrsfs,bbm,amscd}
\usepackage[tbtags]{amsmath}
\usepackage{lastpage}
\usepackage{cases}
\usepackage{subfigure}
\usepackage{epstopdf}
\usepackage{bm}
\usepackage{lineno}
\usepackage[usenames,dvipsnames]{color}

\usepackage[pdftex,colorlinks,linkcolor=black,anchorcolor=black,citecolor=black,urlcolor=black]{hyperref}
\RequirePackage{xspace}
\def\epem       {\ensuremath{e^+e^-}\xspace}
\def\mumu       {\ensuremath{\mu^+\mu^-}\xspace}
\def\tautau     {\ensuremath{\tau^+\tau^-}\xspace}
\def\g          {\ensuremath{\gamma}\xspace}
\def\gaga       {\ensuremath{\gamma\gamma}\xspace}
\mathchardef\Upsilon="7107
\def\Y#1S{\ensuremath{\Upsilon{(#1{\rm S})}}\xspace}
\def\OneS       {\Y1S}
\def\FourS      {\Y4S}
\def\SixS       {\Y6S}

\begin{document}

\hyphenpenalty=5000
\tolerance=1000

\fancyfoot[C]{\small xxxxxx-\thepage}


\title{Measurement of the integrated luminosity of the Phase 2 data of the Belle II experiment}

\maketitle

\newcounter{AffiliationCounter}
\stepcounter{AffiliationCounter}\edef\instBeihang{\protect\theAffiliationCounter}
\stepcounter{AffiliationCounter}\edef\instBNL{\protect\theAffiliationCounter}
\stepcounter{AffiliationCounter}\edef\instBINP{\protect\theAffiliationCounter}
\stepcounter{AffiliationCounter}\edef\instCinvestavIPN{\protect\theAffiliationCounter}
\stepcounter{AffiliationCounter}\edef\instPrague{\protect\theAffiliationCounter}
\stepcounter{AffiliationCounter}\edef\instChiba{\protect\theAffiliationCounter}
\stepcounter{AffiliationCounter}\edef\instConacyt{\protect\theAffiliationCounter}
\stepcounter{AffiliationCounter}\edef\instDESY{\protect\theAffiliationCounter}
\stepcounter{AffiliationCounter}\edef\instDuke{\protect\theAffiliationCounter}
\stepcounter{AffiliationCounter}\edef\instEri{\protect\theAffiliationCounter}
\stepcounter{AffiliationCounter}\edef\instJuelich{\protect\theAffiliationCounter}
\stepcounter{AffiliationCounter}\edef\instFuJen{\protect\theAffiliationCounter}
\stepcounter{AffiliationCounter}\edef\instFudan{\protect\theAffiliationCounter}
\stepcounter{AffiliationCounter}\edef\instGoettingen{\protect\theAffiliationCounter}
\stepcounter{AffiliationCounter}\edef\instGifu{\protect\theAffiliationCounter}
\stepcounter{AffiliationCounter}\edef\instSOKENDAI{\protect\theAffiliationCounter}
\stepcounter{AffiliationCounter}\edef\instGyeongsang{\protect\theAffiliationCounter}
\stepcounter{AffiliationCounter}\edef\instHanyang{\protect\theAffiliationCounter}
\stepcounter{AffiliationCounter}\edef\instKEK{\protect\theAffiliationCounter}
\stepcounter{AffiliationCounter}\edef\instJPARC{\protect\theAffiliationCounter}
\stepcounter{AffiliationCounter}\edef\instIISER{\protect\theAffiliationCounter}
\stepcounter{AffiliationCounter}\edef\instIITBhubaneswar{\protect\theAffiliationCounter}
\stepcounter{AffiliationCounter}\edef\instIITHyderabad{\protect\theAffiliationCounter}
\stepcounter{AffiliationCounter}\edef\instIITMadras{\protect\theAffiliationCounter}
\stepcounter{AffiliationCounter}\edef\instIndiana{\protect\theAffiliationCounter}
\stepcounter{AffiliationCounter}\edef\instIHEPRussia{\protect\theAffiliationCounter}
\stepcounter{AffiliationCounter}\edef\instHEPHYVienna{\protect\theAffiliationCounter}
\stepcounter{AffiliationCounter}\edef\instIHEPChina{\protect\theAffiliationCounter}
\stepcounter{AffiliationCounter}\edef\instIPP{\protect\theAffiliationCounter}
\stepcounter{AffiliationCounter}\edef\instIOP{\protect\theAffiliationCounter}
\stepcounter{AffiliationCounter}\edef\instIFIC{\protect\theAffiliationCounter}
\stepcounter{AffiliationCounter}\edef\instFrascati{\protect\theAffiliationCounter}
\stepcounter{AffiliationCounter}\edef\instNapoliINFN{\protect\theAffiliationCounter}
\stepcounter{AffiliationCounter}\edef\instPadovaINFN{\protect\theAffiliationCounter}
\stepcounter{AffiliationCounter}\edef\instPerugiaINFN{\protect\theAffiliationCounter}
\stepcounter{AffiliationCounter}\edef\instPisaINFN{\protect\theAffiliationCounter}
\stepcounter{AffiliationCounter}\edef\instRomaINFN{\protect\theAffiliationCounter}
\stepcounter{AffiliationCounter}\edef\instRomaTreINFN{\protect\theAffiliationCounter}
\stepcounter{AffiliationCounter}\edef\instTorinoINFN{\protect\theAffiliationCounter}
\stepcounter{AffiliationCounter}\edef\instTriesteINFN{\protect\theAffiliationCounter}
\stepcounter{AffiliationCounter}\edef\instJAEA{\protect\theAffiliationCounter}
\stepcounter{AffiliationCounter}\edef\instMainz{\protect\theAffiliationCounter}
\stepcounter{AffiliationCounter}\edef\instGiessen{\protect\theAffiliationCounter}
\stepcounter{AffiliationCounter}\edef\instKarlsruhe{\protect\theAffiliationCounter}
\stepcounter{AffiliationCounter}\edef\instKitasato{\protect\theAffiliationCounter}
\stepcounter{AffiliationCounter}\edef\instKISTI{\protect\theAffiliationCounter}
\stepcounter{AffiliationCounter}\edef\instKorea{\protect\theAffiliationCounter}
\stepcounter{AffiliationCounter}\edef\instKSU{\protect\theAffiliationCounter}
\stepcounter{AffiliationCounter}\edef\instKyungpook{\protect\theAffiliationCounter}
\stepcounter{AffiliationCounter}\edef\instLAL{\protect\theAffiliationCounter}
\stepcounter{AffiliationCounter}\edef\instLPI{\protect\theAffiliationCounter}
\stepcounter{AffiliationCounter}\edef\instLNNU{\protect\theAffiliationCounter}
\stepcounter{AffiliationCounter}\edef\instLMU{\protect\theAffiliationCounter}
\stepcounter{AffiliationCounter}\edef\instLuther{\protect\theAffiliationCounter}
\stepcounter{AffiliationCounter}\edef\instMNITJaipur{\protect\theAffiliationCounter}
\stepcounter{AffiliationCounter}\edef\instMPP{\protect\theAffiliationCounter}
\stepcounter{AffiliationCounter}\edef\instMPGHLL{\protect\theAffiliationCounter}
\stepcounter{AffiliationCounter}\edef\instMcGill{\protect\theAffiliationCounter}
\stepcounter{AffiliationCounter}\edef\instMIPT{\protect\theAffiliationCounter}
\stepcounter{AffiliationCounter}\edef\instMEPhI{\protect\theAffiliationCounter}
\stepcounter{AffiliationCounter}\edef\instNagoya{\protect\theAffiliationCounter}
\stepcounter{AffiliationCounter}\edef\instNagoyaKMI{\protect\theAffiliationCounter}
\stepcounter{AffiliationCounter}\edef\instNaraWu{\protect\theAffiliationCounter}
\stepcounter{AffiliationCounter}\edef\instNTUTaiwan{\protect\theAffiliationCounter}
\stepcounter{AffiliationCounter}\edef\instKrakow{\protect\theAffiliationCounter}
\stepcounter{AffiliationCounter}\edef\instNiigata{\protect\theAffiliationCounter}
\stepcounter{AffiliationCounter}\edef\instNSU{\protect\theAffiliationCounter}
\stepcounter{AffiliationCounter}\edef\instOkinawa{\protect\theAffiliationCounter}
\stepcounter{AffiliationCounter}\edef\instOsakaCity{\protect\theAffiliationCounter}
\stepcounter{AffiliationCounter}\edef\instRCNP{\protect\theAffiliationCounter}
\stepcounter{AffiliationCounter}\edef\instPNNL{\protect\theAffiliationCounter}
\stepcounter{AffiliationCounter}\edef\instPanjab{\protect\theAffiliationCounter}
\stepcounter{AffiliationCounter}\edef\instPeking{\protect\theAffiliationCounter}
\stepcounter{AffiliationCounter}\edef\instPanjabPAU{\protect\theAffiliationCounter}
\stepcounter{AffiliationCounter}\edef\instRIKEN{\protect\theAffiliationCounter}
\stepcounter{AffiliationCounter}\edef\instSeoul{\protect\theAffiliationCounter}
\stepcounter{AffiliationCounter}\edef\instSPU{\protect\theAffiliationCounter}
\stepcounter{AffiliationCounter}\edef\instSoongsil{\protect\theAffiliationCounter}
\stepcounter{AffiliationCounter}\edef\instLjubljanaJSI{\protect\theAffiliationCounter}
\stepcounter{AffiliationCounter}\edef\instKyiv{\protect\theAffiliationCounter}
\stepcounter{AffiliationCounter}\edef\instTata{\protect\theAffiliationCounter}
\stepcounter{AffiliationCounter}\edef\instTUM{\protect\theAffiliationCounter}
\stepcounter{AffiliationCounter}\edef\instTelAviv{\protect\theAffiliationCounter}
\stepcounter{AffiliationCounter}\edef\instToho{\protect\theAffiliationCounter}
\stepcounter{AffiliationCounter}\edef\instTohoku{\protect\theAffiliationCounter}
\stepcounter{AffiliationCounter}\edef\instTitech{\protect\theAffiliationCounter}
\stepcounter{AffiliationCounter}\edef\instTokyoMetropolitan{\protect\theAffiliationCounter}
\stepcounter{AffiliationCounter}\edef\instUAS{\protect\theAffiliationCounter}
\stepcounter{AffiliationCounter}\edef\instNapoliUNIV{\protect\theAffiliationCounter}
\stepcounter{AffiliationCounter}\edef\instNapoliUNIVA{\protect\theAffiliationCounter}
\stepcounter{AffiliationCounter}\edef\instPadovaUNIV{\protect\theAffiliationCounter}
\stepcounter{AffiliationCounter}\edef\instPerugiaUNIV{\protect\theAffiliationCounter}
\stepcounter{AffiliationCounter}\edef\instPisaUNIV{\protect\theAffiliationCounter}
\stepcounter{AffiliationCounter}\edef\instRomaTreUNIV{\protect\theAffiliationCounter}
\stepcounter{AffiliationCounter}\edef\instTorinoUNIV{\protect\theAffiliationCounter}
\stepcounter{AffiliationCounter}\edef\instTriesteUNIV{\protect\theAffiliationCounter}
\stepcounter{AffiliationCounter}\edef\instMontreal{\protect\theAffiliationCounter}
\stepcounter{AffiliationCounter}\edef\instIPHC{\protect\theAffiliationCounter}
\stepcounter{AffiliationCounter}\edef\instAdelaide{\protect\theAffiliationCounter}
\stepcounter{AffiliationCounter}\edef\instBonn{\protect\theAffiliationCounter}
\stepcounter{AffiliationCounter}\edef\instUBC{\protect\theAffiliationCounter}
\stepcounter{AffiliationCounter}\edef\instCincinnati{\protect\theAffiliationCounter}
\stepcounter{AffiliationCounter}\edef\instFlorida{\protect\theAffiliationCounter}
\stepcounter{AffiliationCounter}\edef\instHawaii{\protect\theAffiliationCounter}
\stepcounter{AffiliationCounter}\edef\instHeidelberg{\protect\theAffiliationCounter}
\stepcounter{AffiliationCounter}\edef\instLjubljanaUniLJ{\protect\theAffiliationCounter}
\stepcounter{AffiliationCounter}\edef\instLouisville{\protect\theAffiliationCounter}
\stepcounter{AffiliationCounter}\edef\instMalaya{\protect\theAffiliationCounter}
\stepcounter{AffiliationCounter}\edef\instLjubljanaUM{\protect\theAffiliationCounter}
\stepcounter{AffiliationCounter}\edef\instMelbourne{\protect\theAffiliationCounter}
\stepcounter{AffiliationCounter}\edef\instMississippi{\protect\theAffiliationCounter}
\stepcounter{AffiliationCounter}\edef\instPittsburgh{\protect\theAffiliationCounter}
\stepcounter{AffiliationCounter}\edef\instUSTC{\protect\theAffiliationCounter}
\stepcounter{AffiliationCounter}\edef\instSAlabama{\protect\theAffiliationCounter}
\stepcounter{AffiliationCounter}\edef\instSCarolina{\protect\theAffiliationCounter}
\stepcounter{AffiliationCounter}\edef\instSydney{\protect\theAffiliationCounter}
\stepcounter{AffiliationCounter}\edef\instUTokyo{\protect\theAffiliationCounter}
\stepcounter{AffiliationCounter}\edef\instIPMU{\protect\theAffiliationCounter}
\stepcounter{AffiliationCounter}\edef\instVictoria{\protect\theAffiliationCounter}
\stepcounter{AffiliationCounter}\edef\instVPI{\protect\theAffiliationCounter}
\stepcounter{AffiliationCounter}\edef\instWayneState{\protect\theAffiliationCounter}
\stepcounter{AffiliationCounter}\edef\instYamagata{\protect\theAffiliationCounter}
\stepcounter{AffiliationCounter}\edef\instYerevan{\protect\theAffiliationCounter}
\stepcounter{AffiliationCounter}\edef\instYonsei{\protect\theAffiliationCounter}
\begin{small}
\begin{center}
F.~Abudin{\'e}n,$^{\instTriesteINFN}$ I.~Adachi,$^{\instKEK,\instSOKENDAI}$ P.~Ahlburg,$^{\instBonn}$ H.~Aihara,$^{\instUTokyo}$ N.~Akopov,$^{\instYerevan}$ A.~Aloisio,$^{\instNapoliUNIV,\instNapoliINFN}$ F.~Ameli,$^{\instRomaINFN}$ L.~Andricek,$^{\instMPGHLL}$ N.~Anh~Ky,$^{\instIOP}$ D.~M.~Asner,$^{\instBNL}$
% ...
\\
\vspace{0.2cm}
(Belle II Collaboration) \\
\vspace{0.2cm}
{\it
$^{\instBeihang}$~Beihang University, Beijing 100191\\
$^{\instBNL}$~Brookhaven National Laboratory, Upton, New York 11973\\
$^{\instBINP}$~Budker Institute of Nuclear Physics SB RAS, Novosibirsk 630090\\
$^{\instCinvestavIPN}$~Centro de Investigacion y de Estudios Avanzados del Instituto Politecnico Nacional, Mexico City 07360\\
$^{\instPrague}$~Faculty of Mathematics and Physics, Charles University, 121 16 Prague\\
$^{\instChiba}$~Chiba University, Chiba 263-8522\\
$^{\instConacyt}$~{Consejo Nacional de Ciencia y Tecnolog\'{\i}a, Mexico City 03940}\\
$^{\instDESY}$~Deutsches Elektronen--Synchrotron, 22607 Hamburg\\
$^{\instDuke}$~Duke University, Durham, North Carolina 27708\\
$^{\instEri}$~Earthquake Research Institute, University of Tokyo, Tokyo 113-0032\\
$^{\instJuelich}$~Forschungszentrum J\"{u}lich, 52425 J\"{u}lich\\
$^{\instFuJen}$~Department of Physics, Fu Jen Catholic University, Taipei 24205\\
$^{\instFudan}$~Key Laboratory of Nuclear Physics and Ion-beam Application (MOE) and Institute of Modern Physics, Fudan University, Shanghai 200443\\
$^{\instGoettingen}$~II. Physikalisches Institut, Georg-August-Universit\"{a}t G\"{o}ttingen, 37073 G\"{o}ttingen\\
$^{\instGifu}$~Gifu University, Gifu 501-1193\\
$^{\instSOKENDAI}$~The Graduate University for Advanced Studies (SOKENDAI), Hayama 240-0193\\
$^{\instGyeongsang}$~Gyeongsang National University, Jinju 52828\\
$^{\instHanyang}$~Department of Physics and Institute of Natural Sciences, Hanyang University, Seoul 04763\\
$^{\instKEK}$~High Energy Accelerator Research Organization (KEK), Tsukuba 305-0801\\
$^{\instJPARC}$~J-PARC Branch, KEK Theory Center, High Energy Accelerator Research Organization (KEK), Tsukuba 305-0801\\
$^{\instIISER}$~Indian Institute of Science Education and Research Mohali, SAS Nagar, 140306\\
$^{\instIITBhubaneswar}$~Indian Institute of Technology Bhubaneswar, Satya Nagar 751007\\
$^{\instIITHyderabad}$~Indian Institute of Technology Hyderabad, Telangana 502285\\
$^{\instIITMadras}$~Indian Institute of Technology Madras, Chennai 600036\\
$^{\instIndiana}$~Indiana University, Bloomington, Indiana 47408\\
$^{\instIHEPRussia}$~Institute for High Energy Physics, Protvino 142281\\
$^{\instHEPHYVienna}$~Institute of High Energy Physics, Vienna 1050\\
$^{\instIHEPChina}$~Institute of High Energy Physics, Chinese Academy of Sciences, Beijing 100049\\
$^{\instIPP}$~Institute of Particle Physics (Canada), Victoria, British Columbia V8W 2Y2\\
$^{\instIOP}$~Institute of Physics, Hanoi\\
$^{\instIFIC}$~Instituto de Fisica Corpuscular, Paterna 46980\\
$^{\instFrascati}$~INFN Laboratori Nazionali di Frascati, I-00044 Frascati\\
$^{\instNapoliINFN}$~INFN Sezione di Napoli, I-80126 Napoli\\
$^{\instPadovaINFN}$~INFN Sezione di Padova, I-35131 Padova\\
$^{\instPerugiaINFN}$~INFN Sezione di Perugia, I-06123 Perugia\\
$^{\instPisaINFN}$~INFN Sezione di Pisa, I-56127 Pisa\\
$^{\instRomaINFN}$~INFN Sezione di Roma, I-00185 Roma\\
$^{\instRomaTreINFN}$~INFN Sezione di Roma Tre, I-00146 Roma\\
$^{\instTorinoINFN}$~INFN Sezione di Torino, I-10125 Torino\\
$^{\instTriesteINFN}$~INFN Sezione di Trieste, I-34127 Trieste\\
$^{\instJAEA}$~Advanced Science Research Center, Japan Atomic Energy Agency, Naka 319-1195\\
$^{\instMainz}$~Johannes Gutenberg-Universit\"{a}t Mainz, Institut f\"{u}r Kernphysik, D-55099 Mainz\\
$^{\instGiessen}$~Justus-Liebig-Universit\"{a}t Gie\ss{}en, 35392 Gie\ss{}en\\
$^{\instKarlsruhe}$~Institut f\"{u}r Experimentelle Teilchenphysik, Karlsruher Institut f\"{u}r Technologie, 76131 Karlsruhe\\
$^{\instKitasato}$~Kitasato University, Sagamihara 252-0373\\
$^{\instKISTI}$~Korea Institute of Science and Technology Information, Daejeon 34141\\
$^{\instKorea}$~Korea University, Seoul 02841\\
$^{\instKSU}$~Kyoto Sangyo University, Kyoto 603-8555\\
$^{\instKyungpook}$~Kyungpook National University, Daegu 41566\\
$^{\instLAL}$~Laboratoire de l'Acc\'{e}l\'{e}rateur Lin\'{e}aire, IN2P3/CNRS et Universit\'{e} Paris-Sud 11, Centre Scientifique d'Orsay, F-91898 Orsay Cedex\\
$^{\instLPI}$~P.N. Lebedev Physical Institute of the Russian Academy of Sciences, Moscow 119991\\
$^{\instLNNU}$~Liaoning Normal University, Dalian 116029\\
$^{\instLMU}$~Ludwig Maximilians University, 80539 Munich\\
$^{\instLuther}$~Luther College, Decorah, Iowa 52101\\
$^{\instMNITJaipur}$~Malaviya National Institute of Technology Jaipur, Jaipur 302017\\
$^{\instMPP}$~Max-Planck-Institut f\"{u}r Physik, 80805 M\"{u}nchen\\
$^{\instMPGHLL}$~Semiconductor Laboratory of the Max Planck Society, 81739 M\"{u}nchen\\
$^{\instMcGill}$~McGill University, Montr\'{e}al, Qu\'{e}bec, H3A 2T8\\
$^{\instMIPT}$~Moscow Institute of Physics and Technology, Moscow Region 141700\\
$^{\instMEPhI}$~Moscow Physical Engineering Institute, Moscow 115409\\
$^{\instNagoya}$~Graduate School of Science, Nagoya University, Nagoya 464-8602\\
$^{\instNagoyaKMI}$~Kobayashi-Maskawa Institute, Nagoya University, Nagoya 464-8602\\
$^{\instNaraWu}$~Nara Women's University, Nara 630-8506\\
$^{\instNTUTaiwan}$~Department of Physics, National Taiwan University, Taipei 10617\\
$^{\instKrakow}$~H. Niewodniczanski Institute of Nuclear Physics, Krakow 31-342\\
$^{\instNiigata}$~Niigata University, Niigata 950-2181\\
$^{\instNSU}$~Novosibirsk State University, Novosibirsk 630090\\
$^{\instOkinawa}$~Okinawa Institute of Science and Technology, Okinawa 904-0495\\
$^{\instOsakaCity}$~Osaka City University, Osaka 558-8585\\
$^{\instRCNP}$~Research Center for Nuclear Physics, Osaka University, Osaka 567-0047\\
$^{\instPNNL}$~Pacific Northwest National Laboratory, Richland, Washington 99352\\
$^{\instPanjab}$~Panjab University, Chandigarh 160014\\
$^{\instPeking}$~Peking University, Beijing 100871\\
$^{\instPanjabPAU}$~Punjab Agricultural University, Ludhiana 141004\\
$^{\instRIKEN}$~Theoretical Research Division, Nishina Center, RIKEN, Saitama 351-0198\\
$^{\instSeoul}$~Seoul National University, Seoul 08826\\
$^{\instSPU}$~Showa Pharmaceutical University, Tokyo 194-8543\\
$^{\instSoongsil}$~Soongsil University, Seoul 06978\\
$^{\instLjubljanaJSI}$~J. Stefan Institute, 1000 Ljubljana\\
$^{\instKyiv}$~Taras Shevchenko National Univ. of Kiev, Kiev\\
$^{\instTata}$~Tata Institute of Fundamental Research, Mumbai 400005\\
$^{\instTUM}$~Department of Physics, Technische Universit\"{a}t M\"{u}nchen, 85748 Garching\\
$^{\instTelAviv}$~Tel Aviv University, School of Physics and Astronomy, Tel Aviv, 69978\\
$^{\instToho}$~Toho University, Funabashi 274-8510\\
$^{\instTohoku}$~Department of Physics, Tohoku University, Sendai 980-8578\\
$^{\instTitech}$~Tokyo Institute of Technology, Tokyo 152-8550\\
$^{\instTokyoMetropolitan}$~Tokyo Metropolitan University, Tokyo 192-0397\\
$^{\instUAS}$~Universidad Autonoma de Sinaloa, Sinaloa 80000\\
$^{\instNapoliUNIV}$~Dipartimento di Scienze Fisiche, Universit\`{a} di Napoli Federico II, I-80126 Napoli\\
$^{\instNapoliUNIVA}$~Dipartimento di Agraria, Universit\`{a} di Napoli Federico II, I-80055 Portici (NA)\\
$^{\instPadovaUNIV}$~Dipartimento di Fisica e Astronomia, Universit\`{a} di Padova, I-35131 Padova\\
$^{\instPerugiaUNIV}$~Dipartimento di Fisica, Universit\`{a} di Perugia, I-06123 Perugia\\
$^{\instPisaUNIV}$~Dipartimento di Fisica, Universit\`{a} di Pisa, I-56127 Pisa\\
$^{\instRomaTreUNIV}$~Dipartimento di Matematica e Fisica, Universit\`{a} di Roma Tre, I-00146 Roma\\
$^{\instTorinoUNIV}$~Dipartimento di Fisica, Universit\`{a} di Torino, I-10125 Torino\\
$^{\instTriesteUNIV}$~Dipartimento di Fisica, Universit\`{a} di Trieste, I-34127 Trieste\\
$^{\instMontreal}$~Universit\'{e} de Montr\'{e}al, Physique des Particules, Montr\'{e}al, Qu\'{e}bec, H3C 3J7\\
$^{\instIPHC}$~Universit\'{e} de Strasbourg, CNRS, IPHC, UMR 7178, 67037 Strasbourg\\
$^{\instAdelaide}$~Department of Physics, University of Adelaide, Adelaide, South Australia 5005\\
$^{\instBonn}$~University of Bonn, 53115 Bonn\\
$^{\instUBC}$~University of British Columbia, Vancouver, British Columbia, V6T 1Z1\\
$^{\instCincinnati}$~University of Cincinnati, Cincinnati, Ohio 45221\\
$^{\instFlorida}$~University of Florida, Gainesville, Florida 32611\\
$^{\instHawaii}$~University of Hawaii, Honolulu, Hawaii 96822\\
$^{\instHeidelberg}$~University of Heidelberg, 68131 Mannheim\\
$^{\instLjubljanaUniLJ}$~Faculty of Mathematics and Physics, University of Ljubljana, 1000 Ljubljana\\
$^{\instLouisville}$~University of Louisville, Louisville, Kentucky 40292\\
$^{\instMalaya}$~National Centre for Particle Physics, University Malaya, 50603 Kuala Lumpur\\
$^{\instLjubljanaUM}$~University of Maribor, 2000 Maribor\\
$^{\instMelbourne}$~School of Physics, University of Melbourne, Victoria 3010\\
$^{\instMississippi}$~University of Mississippi, University, Mississippi 38677\\
$^{\instPittsburgh}$~University of Pittsburgh, Pittsburgh, Pennsylvania 15260\\
$^{\instUSTC}$~University of Science and Technology of China, Hefei 230026\\
$^{\instSAlabama}$~University of South Alabama, Mobile, Alabama 36688\\
$^{\instSCarolina}$~University of South Carolina, Columbia, South Carolina 29208\\
$^{\instSydney}$~School of Physics, University of Sydney, New South Wales 2006\\
$^{\instUTokyo}$~Department of Physics, University of Tokyo, Tokyo 113-0033\\
$^{\instIPMU}$~Kavli Institute for the Physics and Mathematics of the Universe (WPI), University of Tokyo, Kashiwa 277-8583\\
$^{\instVictoria}$~University of Victoria, Victoria, British Columbia, V8W 3P6\\
$^{\instVPI}$~Virginia Polytechnic Institute and State University, Blacksburg, Virginia 24061\\
$^{\instWayneState}$~Wayne State University, Detroit, Michigan 48202\\
$^{\instYamagata}$~Yamagata University, Yamagata 990-8560\\
$^{\instYerevan}$~Alikhanyan National Science Laboratory, Yerevan 0036\\
$^{\instYonsei}$~Yonsei University, Seoul 03722\\
}
\end{center}
\vspace{0.4cm}
\end{small}

\begin{abstract}
From April to July 2018, a data sample at the peak energy of the \FourS resonance was collected with the Belle~II detector at the SuperKEKB electron-positron collider.
This is the first data sample of the Belle~II experiment.
Using Bhabha and digamma events, we measure the integrated luminosity of the data sample to be ($496.3 \pm 0.3 \pm 3.0$)~pb$^{-1}$, where the first uncertainty is statistical and the second  is systematic.
This work provides a basis for future luminosity measurements at Belle~II.
\end{abstract}



\footnotetext[0]{\hspace*{-3mm}\raisebox{0.3ex}{$\scriptstyle\copyright$}2013 Chinese Physical Society and the Institute of High Energy Physics of the Chinese Academy of Sciences and the Institute of Modern Physics of the Chinese Academy of Sciences and IOP Publishing Ltd}

\begin{multicols}{2}

\section{Introduction}
\label{section: Introduction}
Integrated luminosity ($L$) is a basic quantity in high energy physics experiments.
It reflects the size of the data sample, which is crucial to most of the physics studies in collider-based experiments.
It is also the bridge between the number of produced events ($N$) and the cross section ($\sigma$) of any physics process:
\begin{equation}
N = L \cdot \sigma .
\label{equation: Basic formula among number of produced events, integrated luminosity and cross section}
\end{equation}
\noindent According to this relationship, with the integrated luminosity one can calculate the number of produced events from a known cross section or measure the cross section from a determined number of produced events.
The precise measurement of integrated luminosity is thus fundamental to estimating experimental yields accurately and testing theoretical models precisely.

This paper presents a measurement of the integrated luminosity of the first \epem collision data sample collected with the Belle~II detector~\cite{preprint:Belle-II-TDR}. The Belle~II experiment runs at the SuperKEKB accelerator at the High Energy Accelerator Research Organization (KEK) in Tsukuba, Japan.
Belle~II~\cite{preprint:Belle-II-Physics-Book} is a next-generation $B$-factory experiment~\cite{paper:Phys-of-the-B_Fact}.
It is the successor to the Belle experiment~\cite{paper:Phys-of-Belle} and plans to record a dataset of 50 ab$^{-1}$, which is about 50 times the Belle dataset.
With these data, Belle~II aims to search for physics beyond the Standard Model and further study CP violation in the flavor sector, and precisely measure all parameters of the Cabibbo-Kobayashi-Maskawa ``unitarity triangle'' \cite{preprint:Belle-II-Physics-Book}.
The experiment will also study properties of the strong interaction in hadron physics.

Operation of the SuperKEKB accelerator and the Belle~II detector can be divided into three phases: Phase~1, from February to June 2016; Phase~2, from April to July 2018; and Phase~3, from March 2019 onwards.
The data sample under study in this work was recorded during Phase~2.
During this phase, the beams of electrons and positrons collided at the center-of-mass (CM) energy of the \FourS resonance, with a peak instantaneous luminosity of $5.55 \times 10^{33}$~cm$^{-2}$s$^{-1}$, and the data sample was collected with a nearly complete Belle~II detector. (The full vertex detector was not yet installed; see the next section for the detector description.)
In the earlier Phase~1, the beams were circulated but not collided in the accelerator's storage rings for beam-line conditioning, accelerator performance tuning, and beam background studies~\cite{paper:Phase-1}.
In current and future Phase~3 running, copious data samples of beam-collision events are recorded for the comprehensive physics program of Belle~II. The luminosity measurement of the collision data in Phase~2 is necessary for physics measurements with this data, and is valuable preparation for future measurements in Phase~3.

In \epem collision experiments, the integrated luminosity is mainly measured according to Eq.~(\ref{equation: Basic formula among number of produced events, integrated luminosity and cross section}) with the following two well-known quantum electrodynamics processes: Bhabha scattering \epem $\to$ \epem (n\g) and digamma production \epem $\to$ \gaga (n\g)~\cite{paper:Belle-Lum,paper:BaBar-Lum,paper:BESIII-Lum-1,paper:BESIII-Lum-2,paper:BESIII-Lum-3,paper:BESIII-Lum-4}.
Here, n\g in the Bhabha process involves both the initial-state and final-state radiation photons, while n\g in the digamma process only refers to the initial-state radiation photons.
These two processes have large production rates, accurate theoretical predictions for the cross sections, and simple event topologies that can be simulated precisely and selected with essentially no background contamination.
These three features reduce the statistical and systematic uncertainties, making the Bhabha and digamma processes ideal for integrated luminosity measurements.
In this work, we perform two independent measurements with these two processes; the separate measurements cross-check our methodology.

\section{The Belle II detector}
\label{section: The Belle II detector}
The Belle~II detector records the signals of the final state particles produced in \epem collisions to study the decays of $B$ mesons, charmed particles, $\tau$ leptons, and $\Upsilon(n{\rm S})$ ($n=1,2\cdots6$) resonances as well as the production of new states of matter.
It operates at the SuperKEKB accelerator, which is the upgraded version of the KEKB accelerator, a 3-km-circumference asymmetric-energy electron-positron collider with two storage rings: one for the electron beam, and the other for the positron beam.
The two beams in SuperKEKB collide at a crossing angle of 83 mrad, larger than the crossing angle of 22 mrad in KEKB.
Similar to KEKB, SuperKEKB is designed to work in the energy region from \OneS to \SixS and to operate mainly at the \FourS.
The instantaneous luminosity goal of SuperKEKB is $8 \times 10^{35}$~cm$^{-2}$s$^{-1}$, which is about 40 times higher than that of KEKB.
Notably, due to the asymmetric energies and acollinear orbits of the electron and positron beams, the coordinate system of the laboratory frame is significantly different from that of the CM frame. In particular, in the laboratory frame the $z$ axis is along the bisector of the angle between the direction of the electron beam and the reverse direction of the positron beam, while in the CM frame the $z$ axis is along the direction of the electron beam. Specifically, the $z$ axis in the CM frame points at the same direction as the unit vector (0.1505, 0, 0.9886) in the laboratory frame.

The Belle~II detector surrounds the interaction point (IP), which is within a 1-cm radius beam pipe.
It has a cylindrical structure aligned centrally to the $z$ axis in the laboratory frame and consists of several nested sub-detectors and a superconducting solenoidal magnet.
Six layers of vertex detectors (VXD), including two inner layers of silicon pixel detectors surrounded by four layers of silicon strip detectors, are designed to accurately reconstruct the decay vertices of $B$ mesons and other short-lived particles.
During Phase~2, only a small fraction of the VXD sensors were installed for diagnostic purposes, and the remainder of the VXD volume was instrumented with specialized radiation detectors for beam background measurements~\cite{paper:Beam-Bkgs}.
A small-cell, helium-based (50\% He, 50\% ${\rm C}_2{\rm H}_6$) central drift chamber (CDC) is used to precisely measure the trajectories, momenta, and ionization energy losses of charged  particles.
A particle identification system, including an imaging time-of-propagation (TOP) detector in the barrel region and an aerogel ring imaging Cherenkov detector in the forward endcap region, is used to identify  charged particles.
An electromagnetic calorimeter (ECL), composed of 8736 CsI(Tl) crystals arranged in a barrel and two endcaps, detects photons and provides discrimination of electrons from hadrons --- in particular, pions.
The closely-packed crystals are designed with a tower structure pointing to the IP, but are tilted by $2.5^{\circ}$ in $\theta$ and $\phi$ from the radial line to the IP to avoid the possibility that a photon (or electron) could travel along an inter-crystal gap without showering.
A superconducting solenoid magnet provides a 1.5~T magnetic field for the measurement of the momenta of charged  particles.
The $K_L^0$ and muon detector is a ``sandwich'' of alternating layers of 4.7-cm-thick iron plates and 4.4-cm-thick active detector elements. The latter consists of scintillator strips read out by silicon photomultipliers in the endcap and innermost barrel layers, and glass-electrode resistive plate chambers in the outer barrel layers. This detector is used for the identification of high momentum muons and the detection of $K_L^0$ mesons.
The Belle~II detector is described in detail elsewhere~\cite{preprint:Belle-II-TDR}.

In Bhabha and digamma events, the final-state particles are electrons, positrons, and photons; thus the sub-detectors most vital for the measurements are the VXD, CDC, and ECL.
Since the VXD acceptance was quite limited and the CDC tracking efficiency was relatively low in Phase~2, luminosity measurements using ECL information alone are presented in this paper.
To avoid the uninstrumented gaps between the ECL barrel and endcap regions where the material model in the Monte Carlo (MC) simulation was not well-defined, only information from the ECL barrel region is used in the measurements.

\section{Monte Carlo simulation}
\label{section: Monte Carlo simulation}
To determine detection efficiencies, five million  Bhabha events and one million  digamma events were simulated at the peak energy of the \FourS resonance with

\end{multicols}

\begin{figure}[!h]
\centering
\subfigure{\includegraphics[width=0.375\textwidth]{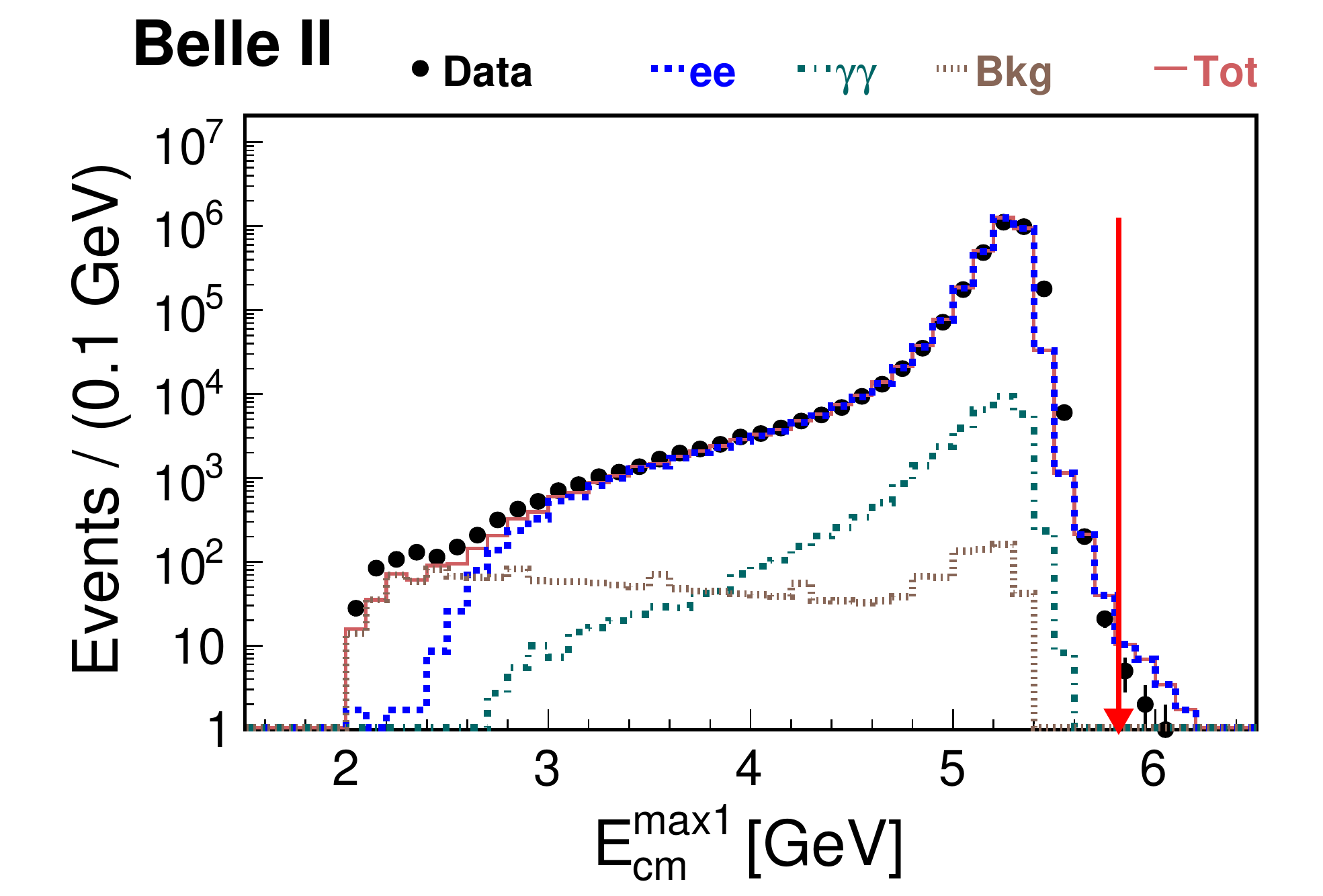}}
\subfigure{\includegraphics[width=0.375\textwidth]{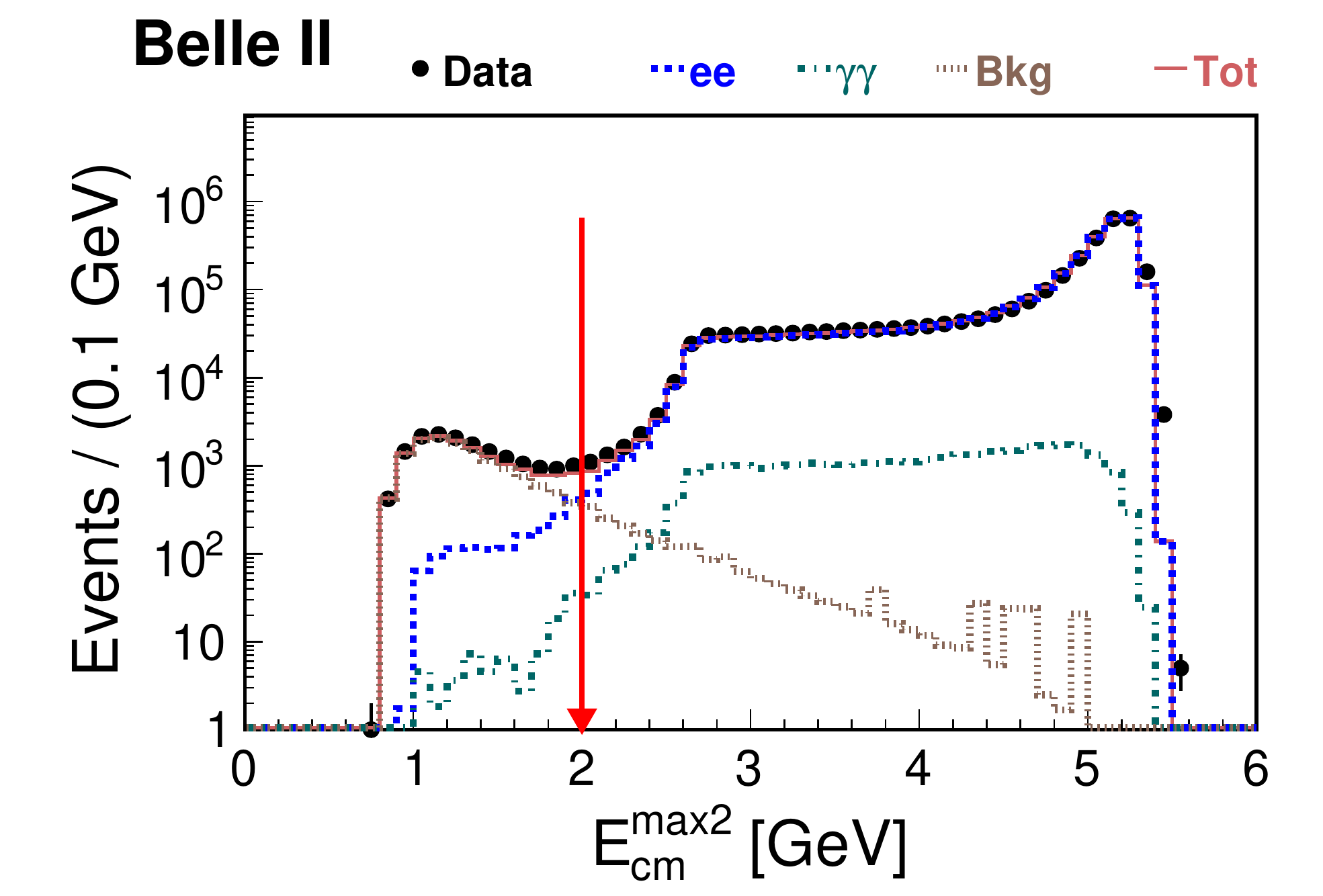}}
\subfigure{\includegraphics[width=0.375\textwidth]{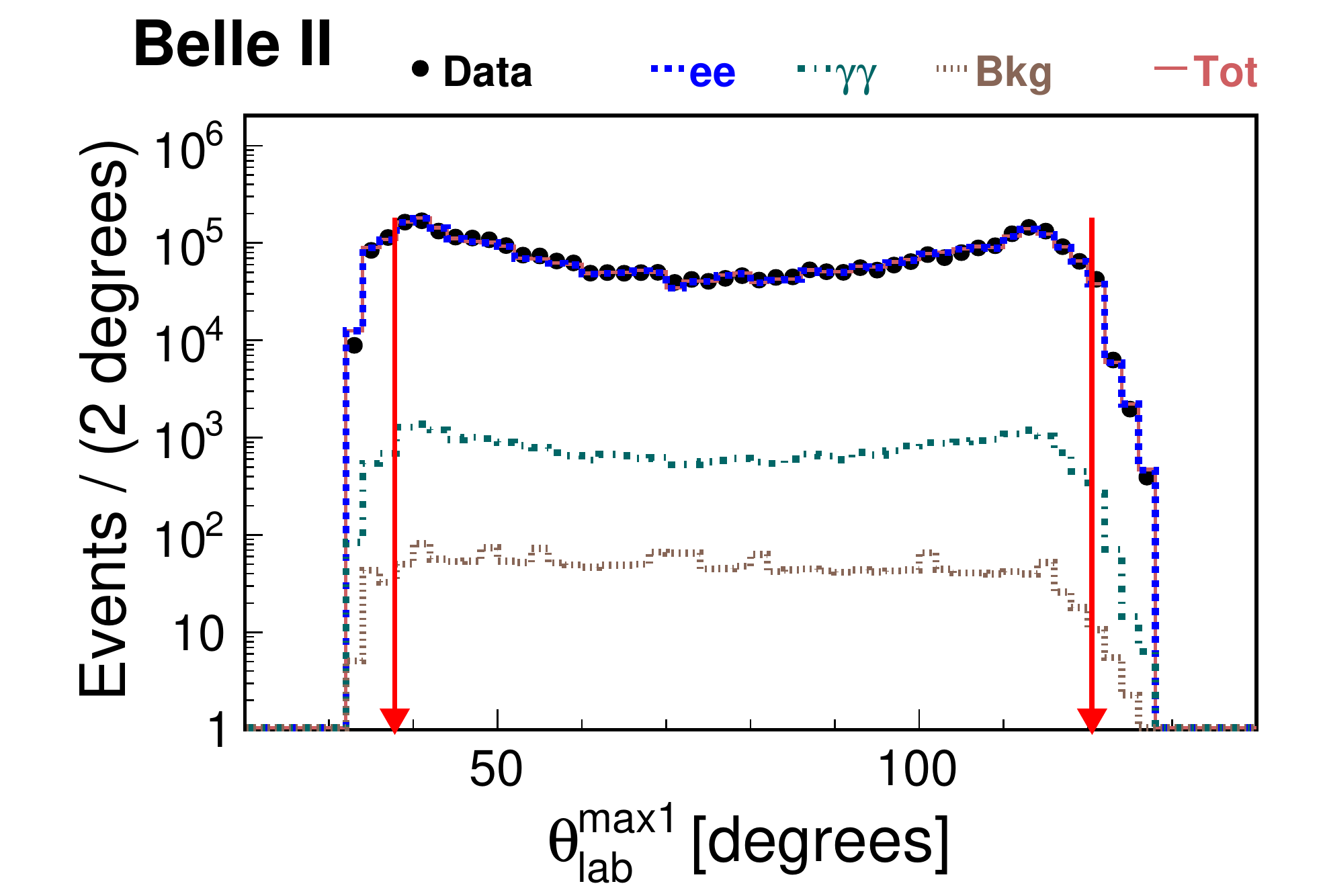}}
\subfigure{\includegraphics[width=0.375\textwidth]{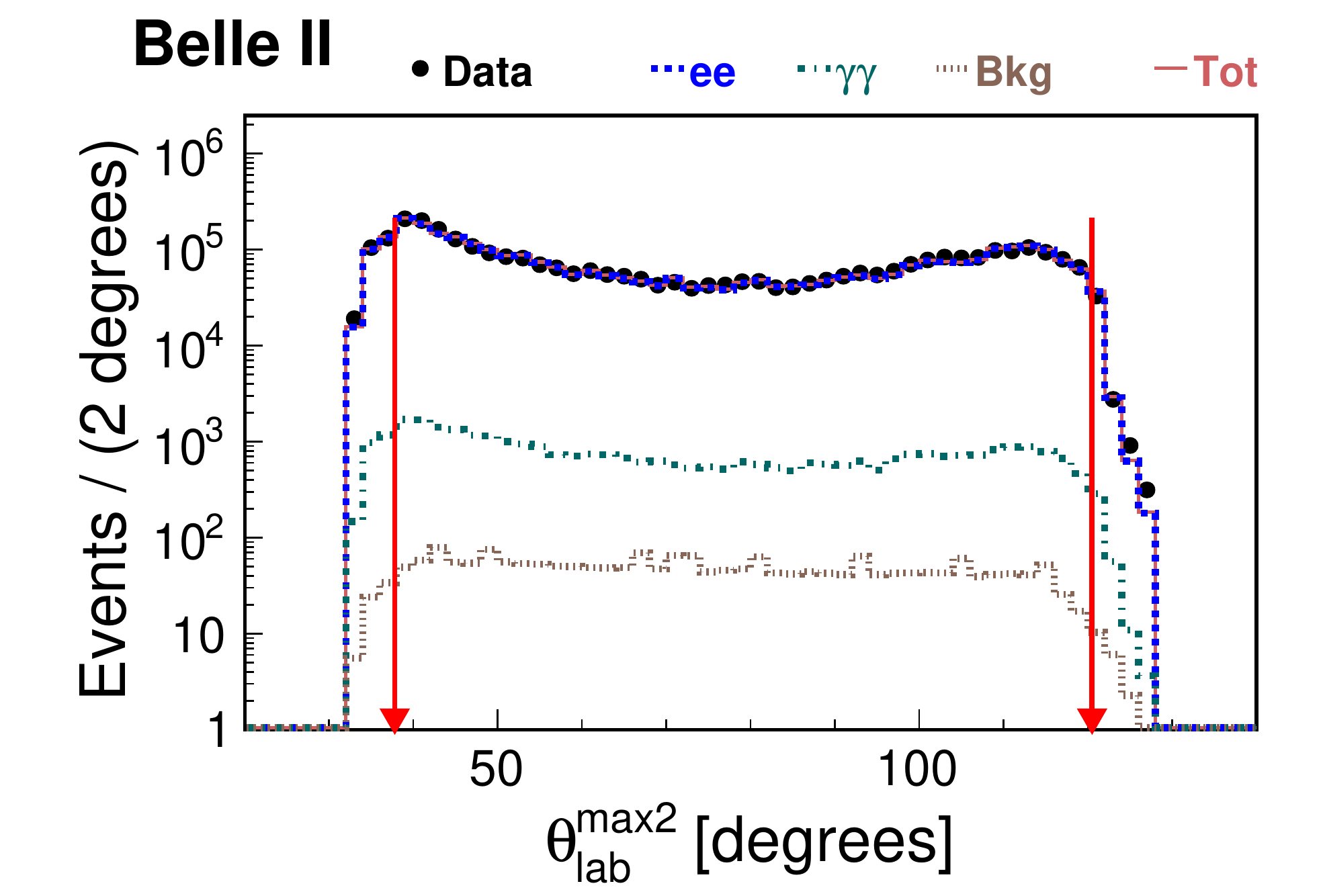}}
\subfigure{\includegraphics[width=0.375\textwidth]{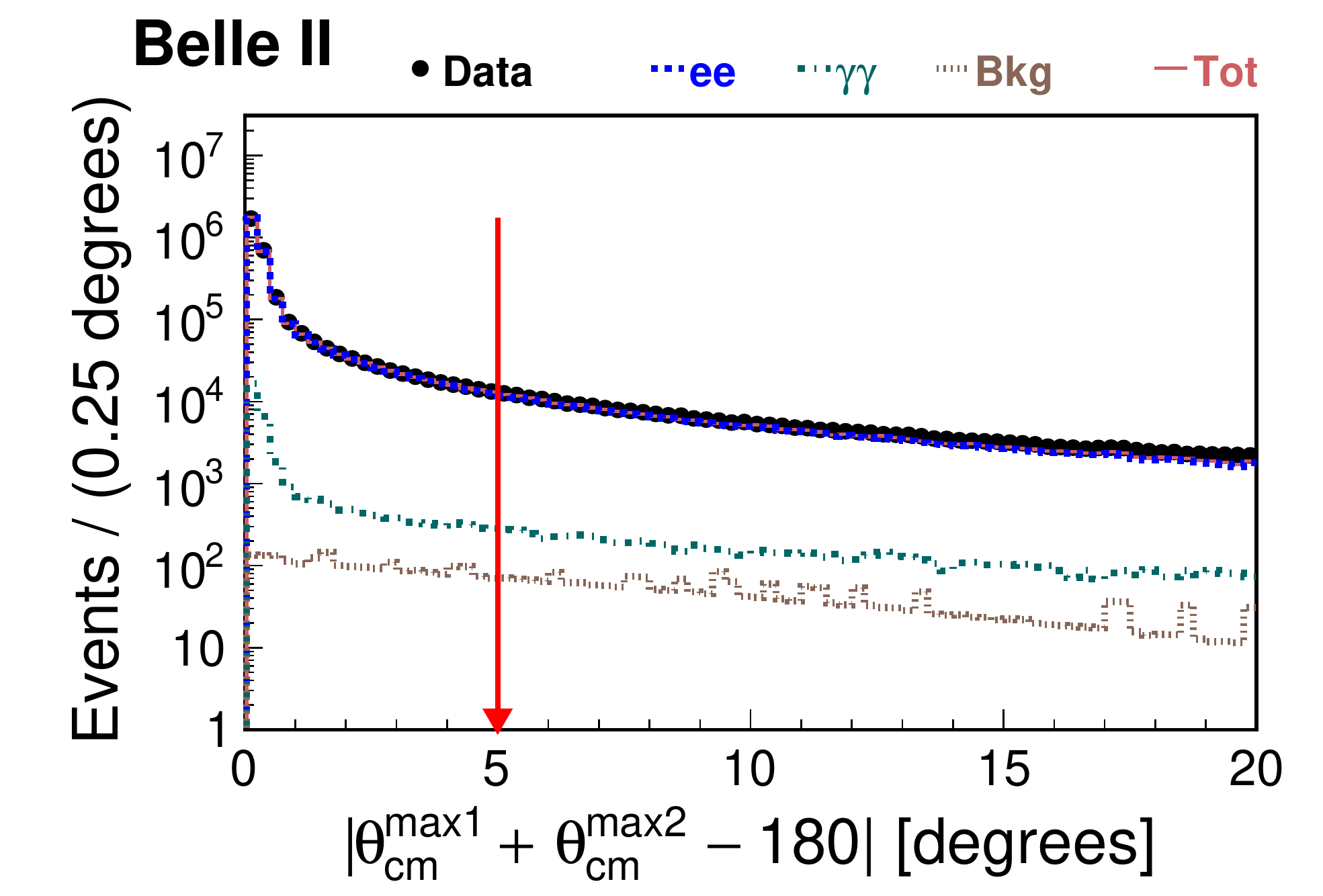}}
\subfigure{\includegraphics[width=0.375\textwidth]{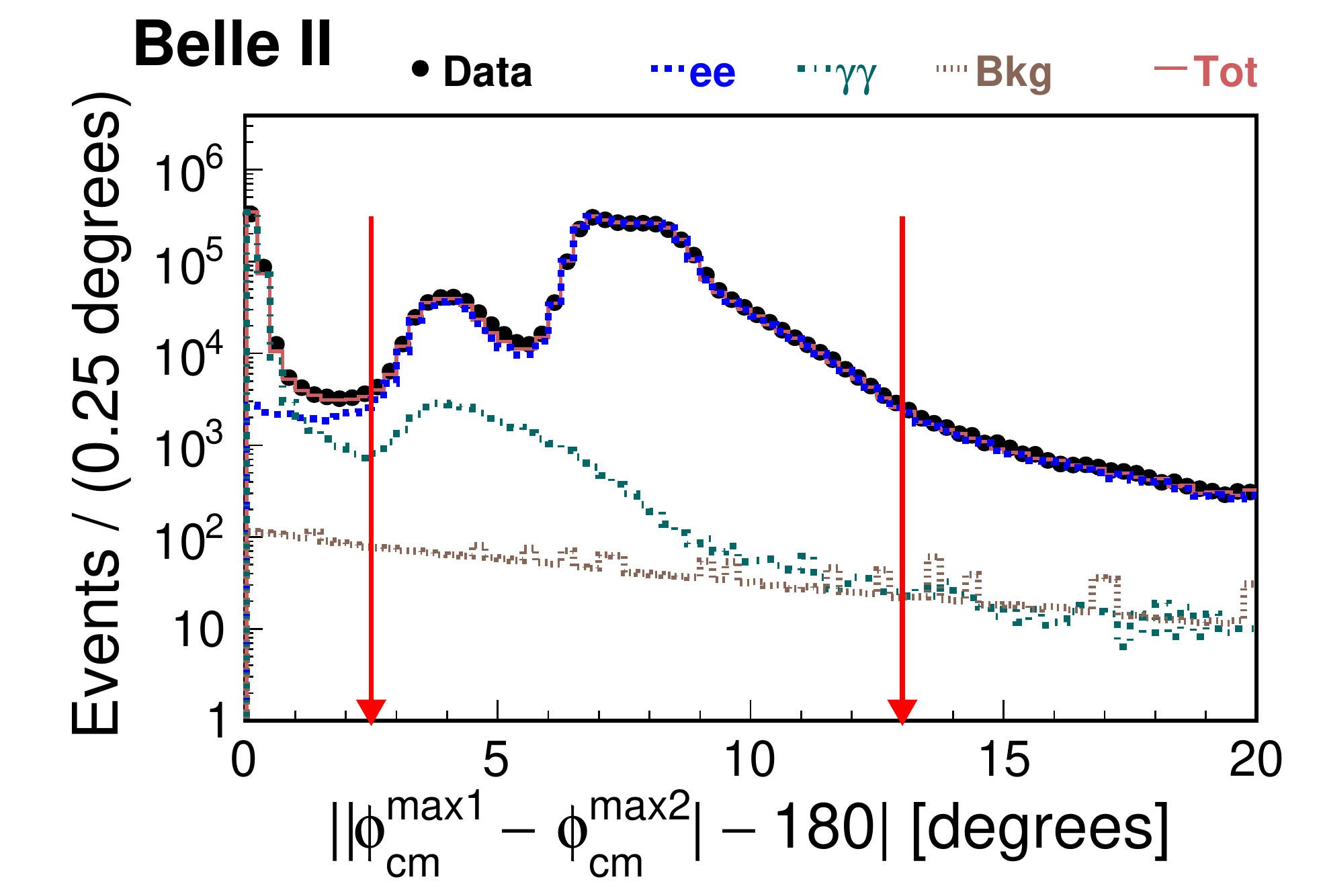}}
\caption{Comparisons of the distributions of Bhabha-dominant signal candidates between the data and MC samples. Each plot in the figure shows one quantity in the selection criteria and is drawn with the requirements on all other quantities applied. In the legend, ``Data'' represents the data sample, while ``ee'', ``$\gamma\gamma$'', ``Bkg'', and ``Tot'' denote the Bhabha, digamma, background (\mumu, $e^{+}e^{-}e^{+}e^{-}$, $B^{+}B^{-}$, $B^0\bar{B^0}$, $c\bar{c}$, $s\bar{s}$, $u\bar{u}$, $d\bar{d}$, and \tautau), and total MC samples, respectively. The vertical arrows indicate the regions of the selected events.}
\label{figure: Comparisons of the distributions of Bhabha candidates between the data and MC samples.}
\end{figure}

\begin{figure}[!h]
\centering
\subfigure{\includegraphics[width=0.375\textwidth]{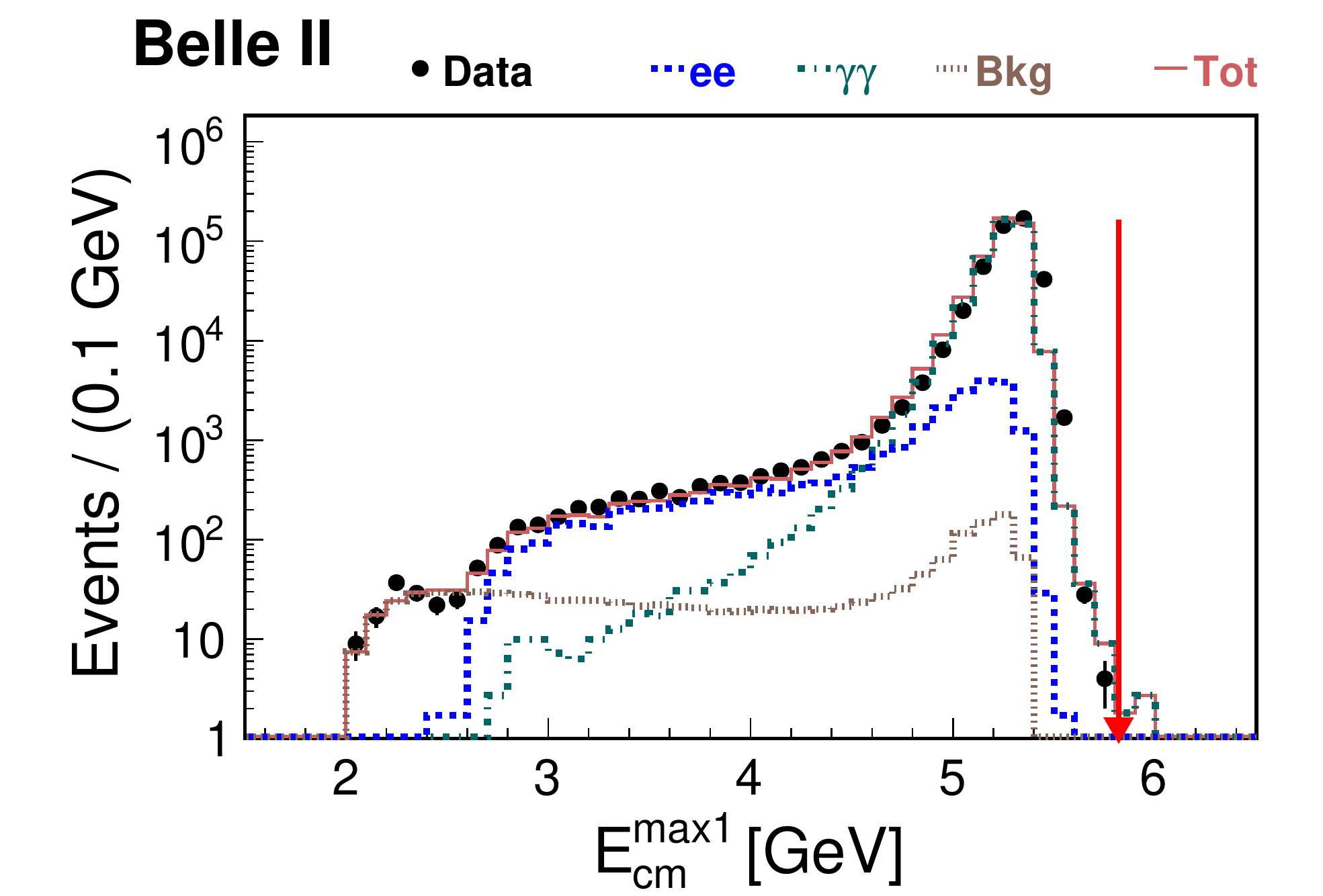}}
\subfigure{\includegraphics[width=0.375\textwidth]{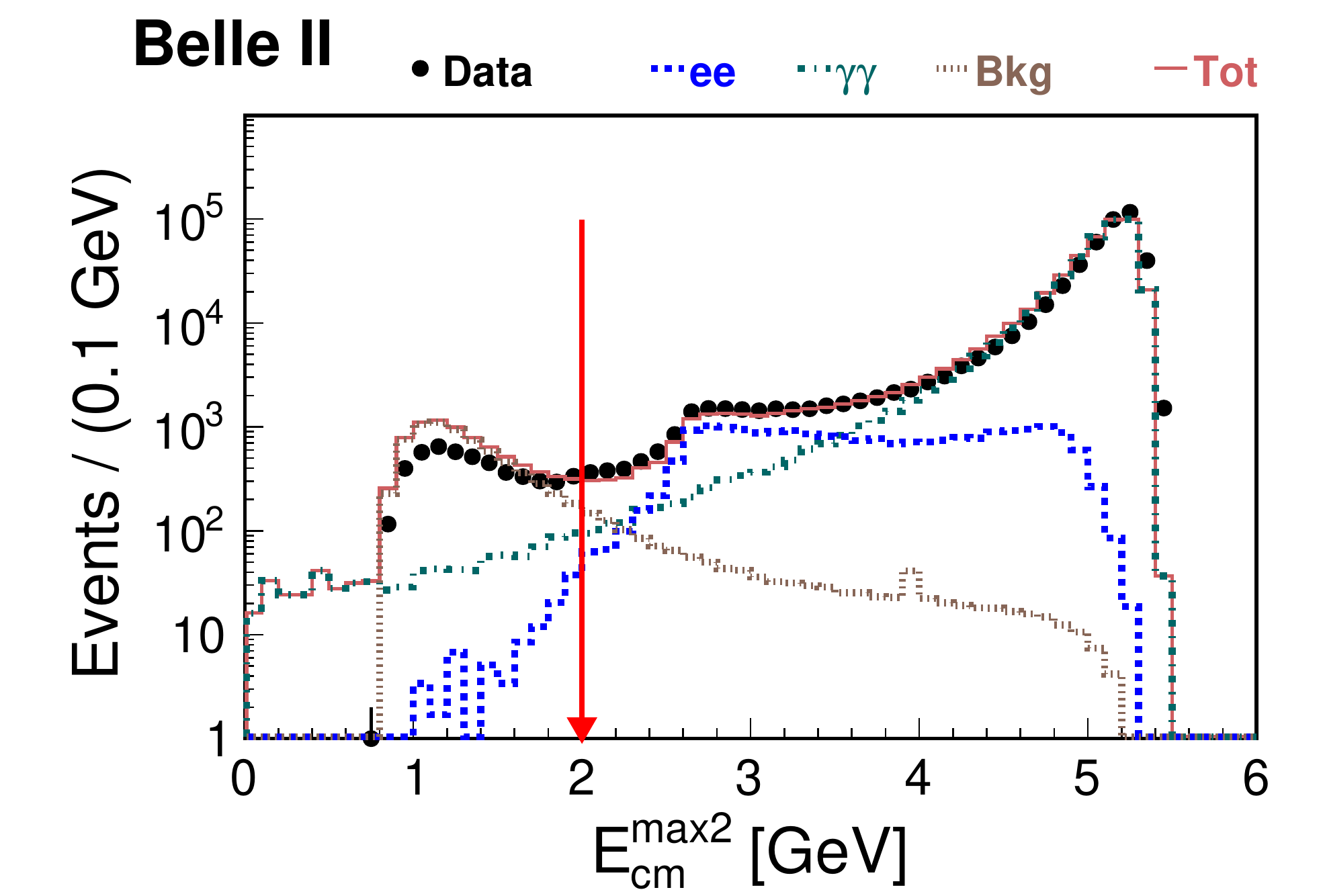}}
\subfigure{\includegraphics[width=0.375\textwidth]{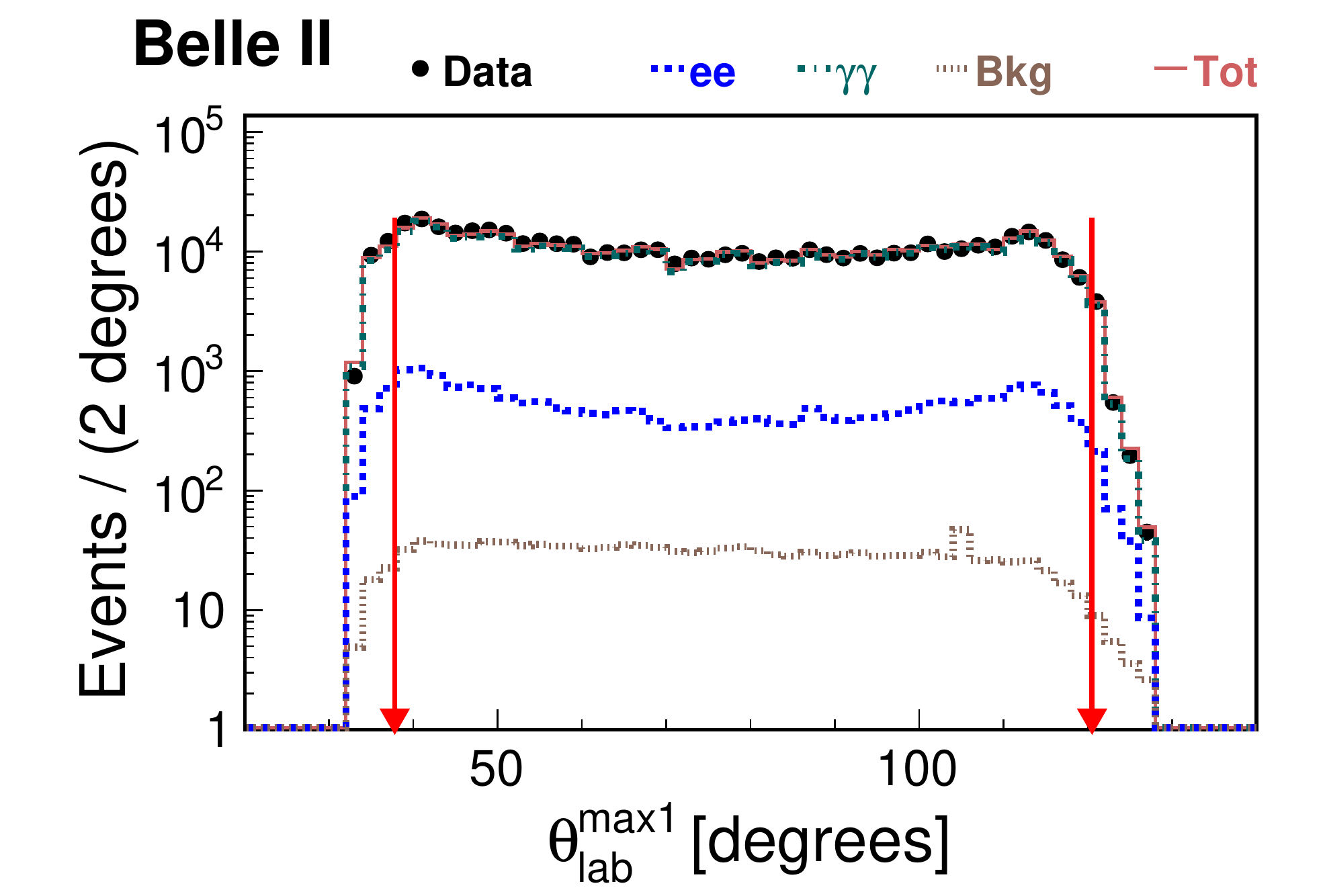}}
\subfigure{\includegraphics[width=0.375\textwidth]{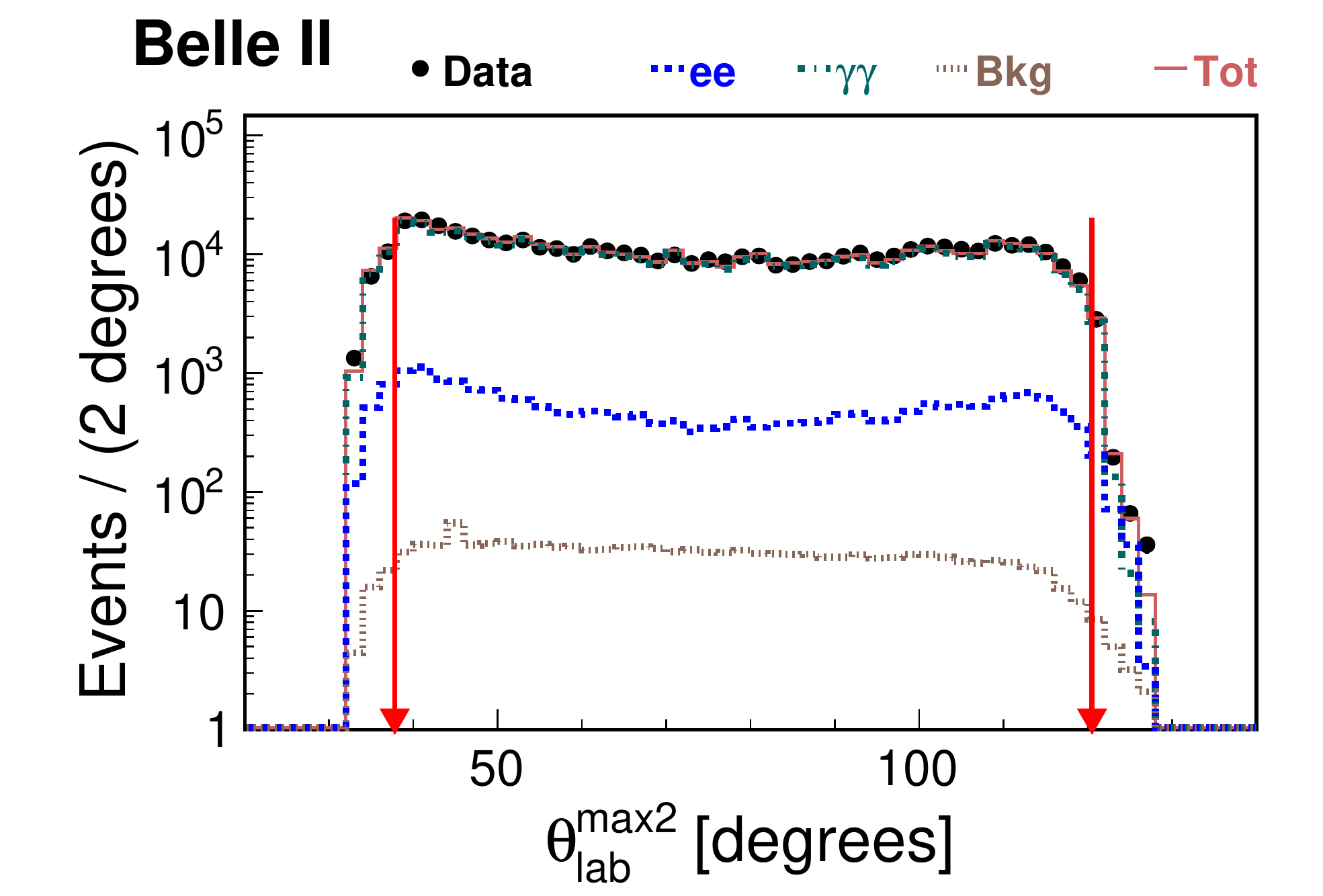}}
\subfigure{\includegraphics[width=0.375\textwidth]{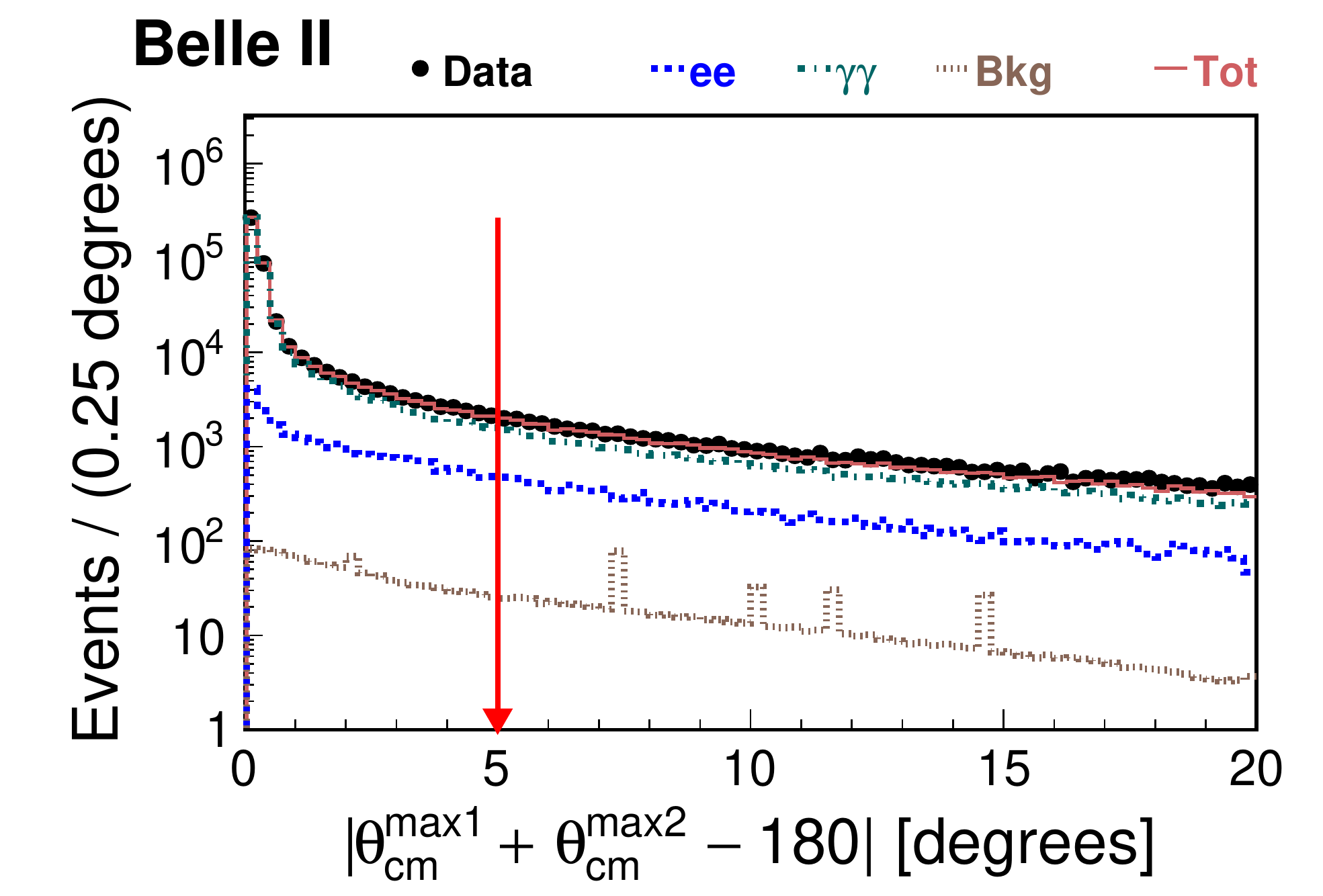}}
\subfigure{\includegraphics[width=0.375\textwidth]{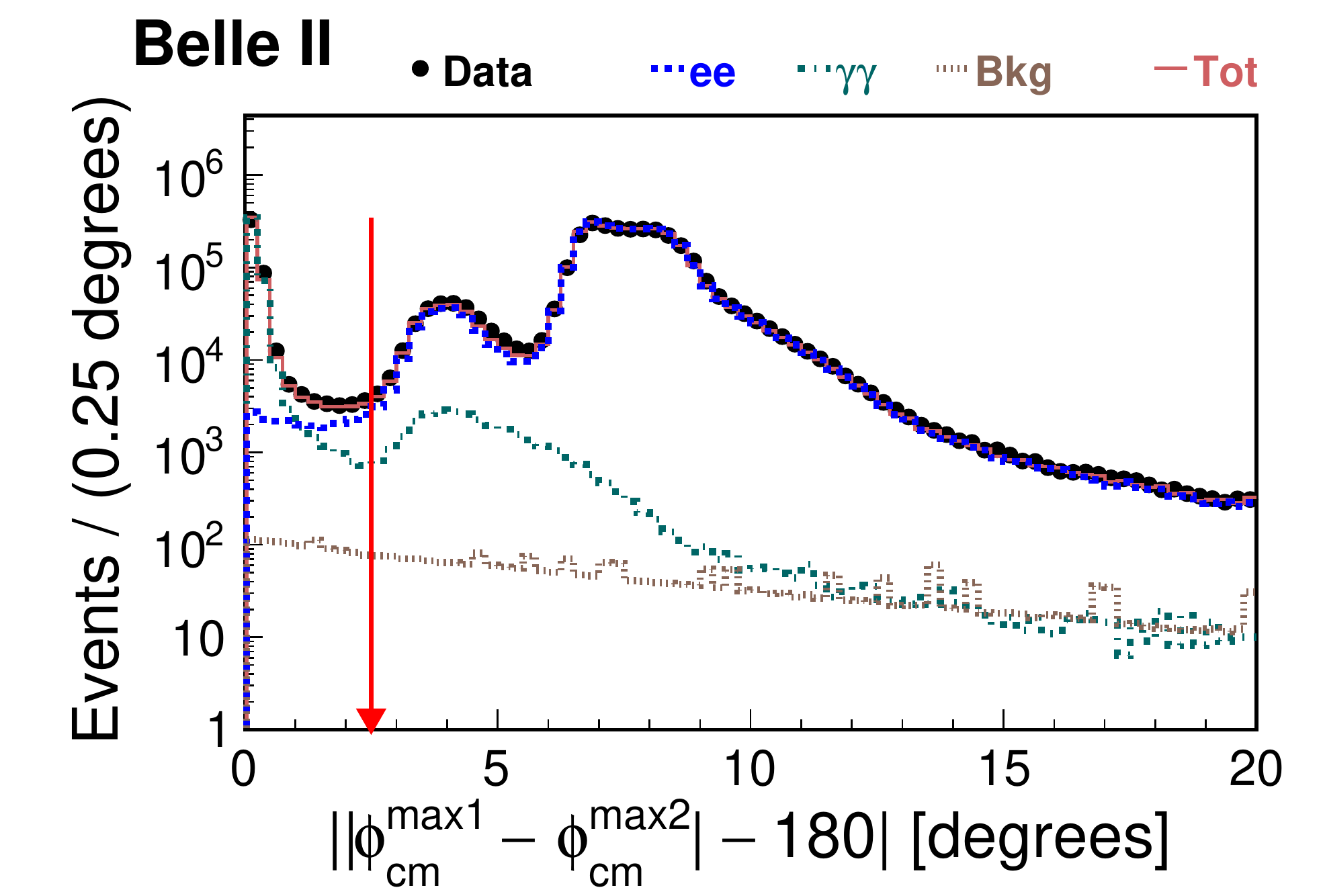}}
\caption{Comparisons of the distributions of digamma-dominant signal candidates between the data and MC samples. Each plot in the figure shows one quantity in the selection criteria and is drawn with the requirements on all other quantities applied. In the legend, ``Data'' represents the data sample, while ``ee'', ``$\gamma\gamma$'', ``Bkg'', and ``Tot'' denote the Bhabha, digamma, background (\mumu, $e^{+}e^{-}e^{+}e^{-}$, $B^{+}B^{-}$, $B^0\bar{B^0}$, $c\bar{c}$, $s\bar{s}$, $u\bar{u}$, $d\bar{d}$, and \tautau), and total MC samples, respectively. The vertical arrows indicate the regions of the selected events.}
\label{figure: Comparisons of the distributions of Digamma candidates between the data and MC samples.}
\end{figure}

\begin{multicols}{2}

\noindent a CM beam energy spread of 5~MeV~\cite{Akai:2018mbz} using the {\sc BabaYaga@NLO}~\cite{paper:BabaYaga@NLO-1,paper:BabaYaga@NLO-2,paper:BabaYaga@NLO-3,paper:BabaYaga@NLO-4} generator.
The MC samples were generated in the polar angle range 35$^{\circ}$--145$^{\circ}$ in the CM frame, somewhat broader than the acceptance of the ECL barrel region, to avoid spurious edge effects.
Along with the generation of the samples, the theoretical cross sections of Bhabha and digamma processes ($\sigma_{\rm ee}$ and $\sigma_{\rm \gamma\gamma}$) were evaluated using the same generator with the same input parameters.
The cross sections were calculated to be $\sigma_{\rm ee}=17.37$~nb and $\sigma_{\rm \gamma\gamma}=1.833$~nb with a claimed precision of 0.1\%~\cite{paper:BabaYaga@NLO-1,paper:BabaYaga@NLO-2,paper:BabaYaga@NLO-3,paper:BabaYaga@NLO-4}.

To estimate background levels, the following MC samples were also produced at the peak energy of the \FourS resonance:
one million \mumu events with the {\sc BabaYaga@NLO} generator;
one million  two-photon events in the $e^{+}e^{-}e^{+}e^{-}$ final state with the {\sc AAFH}~\cite{paper:AAFH-1,paper:AAFH-2,paper:AAFH-3} generator;
50-fb$^{-1}$-equivalent of $B^{+}B^{-}$ and $B^0\bar{B^0}$

\noindent events decayed with {\sc EvtGen} 1.3~\cite{paper:EvtGen} for exclusive modes and {\sc PYTHIA} 8.2~\cite{paper:PYTHIA} for inclusive modes;
50-fb$^{-1}$-equivalent of $c\bar{c}$, $s\bar{s}$, $u\bar{u}$, and $d\bar{d}$ events produced with {\sc KKMC} 4.15~\cite{paper:KKMC-1,paper:KKMC-2} and decayed with {\sc EvtGen} 1.3 and {\sc PYTHIA} 8.2;
and 50-fb$^{-1}$-equivalent of \tautau events also produced with {\sc KKMC} 4.15 but decayed with {\sc TAUOLA}~\cite{paper:TAUOLA}.

In order to simulate the interaction of final-state particles with the detector, the generated MC samples were used as input for a {\sc Geant4}-based MC simulation program~\cite{paper:Geant4}, which includes the geometric description and response of the detector.
In the simulation, beam backgrounds, such as those arising from the Touschek effect and beam-gas interactions, were overlaid on the $e^+e^-$ collision events.
The beam backgrounds were first simulated with dedicated accelerator-design software~\cite{paper:SAD}, and then processed by {\sc Geant4} to handle the interactions of the primary beam-background particles with the accelerator and detector material.~\cite{paper:Accelerator design at SuperKEKB}.
Notably, a complete simulation of the material in the VXD region, including the cables, electronics, and support structure, was not yet available at this early stage of the experiment.
The unsimulated material is conservatively estimated to be 20\% of the simulated material, and its impact on the measured luminosities is discussed in Section \ref{section: Systematic uncertainties}.

Both the data and MC samples were reconstructed and analyzed with the Belle~II analysis software framework, basf2~\cite{paper:Basf2}.

\section{Event selection}
\label{section: Event selection}
To determine the integrated luminosity of the data sample, we first select the signals, namely Bhabha and digamma events.
For this purpose, we require that candidate events have at least two ECL clusters, and we subsequently identify the two clusters with the largest energies in the CM frame.
Because the ECL energies for the electrons and positrons of Bhabha events, and the photons of digamma events, tend to be distributed near half the CM energy, the higher energy of the two clusters in the CM frame is required to be less than 5.82~GeV, and the lower energy of the pair is required to be greater than 2~GeV.
To guarantee that the two clusters are well reconstructed within the ECL barrel region, their polar angles, i.e. those of the position vectors of the cluster centers (similar definition applies to their azimuthal angles), in the laboratory frame are required to be in the range 37.8$^{\circ}$--120.5$^{\circ}$.
Since the final-state particles in Bhabha and digamma events are back to back, the acollinearity of the two clusters in polar angle, namely the absolute difference between 180$^{\circ}$ and the sum of the two polar angles in the CM frame, is required to be less than 5$^{\circ}$.
Because of the deflection of electrons and positrons in the magnetic field, the acollinearity of the two clusters in azimuthal angle, namely the absolute difference between 180$^{\circ}$ and the absolute difference of the two azimuthal angles in the CM frame, which peaks around 8$^{\circ}$, is required to be in the range 2.5$^{\circ}$--13$^{\circ}$ for Bhabha events.
Since photons are not affected by the magnetic field, the acollinearity in  azimuthal angle in the CM frame is required to be less than 2.5$^{\circ}$ to select digamma events.
Because the requirements on azimuthal acollinearity do not fully separate Bhabha and digamma events, we refer to the measurements made from each selection as the Bhabha-dominant or digamma-dominant, respectively.
In summary, the selection criteria are as follows.
The common requirements for the two measurements are
\begin{itemize}
\item 2~GeV $<{\rm E}_{\rm cm}^{\rm max2}<{\rm E}_{\rm cm}^{\rm max1}<$ 5.82~GeV,
\item $37.8^{\circ}<\theta_{\rm lab}^{{\rm max}1}$, $\theta_{\rm lab}^{{\rm max}2}<120.5^{\circ}$, and
\item $|\theta_{\rm cm}^{\rm max1}+\theta_{\rm cm}^{\rm max2}-180^\circ|<5^{\circ}$.
\end{itemize}
\noindent Bhabha-dominant events are further selected with
\begin{itemize}
\item $2.5^{\circ}<||\phi_{\rm cm}^{\rm max1}-\phi_{\rm cm}^{\rm max2}|-180^\circ|<13^{\circ}$,
\end{itemize}
\noindent and digamma-dominant events are further selected with
\begin{itemize}
\item $||\phi_{\rm cm}^{\rm max1}-\phi_{\rm cm}^{\rm max2}|-180^\circ|<2.5^{\circ}$.
\end{itemize}

\noindent Here, ${\rm E}$, $\theta$, and $\phi$ denote the energy, polar angle, and azimuthal angle of a cluster.
The subscript cm (lab) denotes the CM (laboratory) frame, and the superscript max1 (max2) identifies the cluster with the largest (second-largest) energy.

The criteria presented above are chosen on the basis of the distributions in Figs.~\ref{figure: Comparisons of the distributions of Bhabha candidates between the data and MC samples.} and \ref{figure: Comparisons of the distributions of Digamma candidates between the data and MC samples.}, which demonstrate the close agreement of the distributions between the data and MC samples for Bhabha-dominant and digamma-dominant measurements, respectively.
Each plot in the figures shows one quantity in the selection criteria and is drawn with the requirements on all other quantities applied.
For example, the top-left plot in Fig.~\ref{figure: Comparisons of the distributions of Bhabha candidates between the data and MC samples.} shows the ${\rm E}_{\rm cm}^{\rm max1}$ distribution for events that satisfy the requirements on ${\rm E}_{\rm cm}^{\rm max2}$, $\theta_{\rm lab}^{{\rm max}1}$, $\theta_{\rm lab}^{{\rm max}2}$, $|\theta_{\rm cm}^{\rm max1}+\theta_{\rm cm}^{\rm max2}-180^\circ|$, and $||\phi_{\rm cm}^{\rm max1}-\phi_{\rm cm}^{\rm max2}|-180^\circ|$.
In the figures, the luminosities of the MC samples are first normalized to a common reference luminosity and then normalized as a whole to the number of events in the data sample in each plot.

In the figures, one sees that the data and MC samples agree quite well except in the following cases.
In the ${\rm E}_{\rm cm}^{\rm max1}$ and ${\rm E}_{\rm cm}^{\rm max2}$ plots, data and MC disagree around the peaks due to the imperfect ECL calibration at this early stage of the experiment.
However, this has a negligible impact on our measurements, because the selection requirements on ${\rm E}_{\rm cm}^{\rm max1}$ and ${\rm E}_{\rm cm}^{\rm max2}$ are far from the peaks.
In addition, we note that the peak around 4$^\circ$ in the $||\phi_{\rm cm}^{\rm max1}-\phi_{\rm cm}^{\rm max2}|-180^\circ|$ plots is mainly associated with Bhabha events with hard final state radiation where the photon, which is not deflected in the magnetic field, has a higher energy than the electron or positron from which it is radiated.
Due to the gamma-conversion effect, digamma events also contribute to this peak, but at a level one order of magnitude smaller.

\section{Determination of the luminosity}
\label{section: Determination of the luminosity}
\begin{center}\tabcaption{\label{table: Measured integrated luminosities and the quantities used to calculate them.} Measured integrated luminosities and the quantities used to calculate them. The second and third columns list the quantities in the Bhabha-dominant and digamma-dominant measurements, respectively. The uncertainties are statistical only.}
\footnotesize
\begin{tabular}{ccc}
\toprule
Quantity & Bhabha & digamma \\
\hline
$N_{\rm data}^{\rm obs}$ & $3\,134\,488\pm1\,770$ & $454\,650\pm674$ \\
$\epsilon_{\rm ee}$ (\%) & $35.93\pm0.02$ & $0.255\pm0.002$ \\
$\epsilon_{\rm \gamma\gamma}$ (\%) & $3.56\pm0.02$ & $47.74\pm0.05$ \\
$\sigma_{\rm ee}$ (nb) & 17.37 & 17.37\\
$\sigma_{\rm \gamma\gamma}$ (nb) &  1.833 & 1.833 \\
$R_{\rm bkg}$ (\%) & 0.07 & 0.28 \\
\hline
$L$ (pb$^{-1}$) & $496.7 \pm 0.3$ & $493.1 \pm 0.7$ \\
\bottomrule
\end{tabular}
\end{center}

In both of the Bhabha-dominant and digamma-dominant measurements, with their respective selection criteria applied, we obtain the number of candidate events ($N_{\rm data}^{\rm obs}$) observed in the data sample, and the detection efficiencies of Bhabha and digamma events ($\epsilon_{\rm ee}$ and $\epsilon_{\rm \gamma\gamma}$) estimated using their respective MC samples, as listed in Table~\ref{table: Measured integrated luminosities and the quantities used to calculate them.}.
Similarly, all the residual efficiencies of the individual categories of backgrounds ($\epsilon_{\rm bkg}$) are estimated with their corresponding MC samples.

Combining the selection efficiencies with the theoretical cross sections of the signal processes as well as those of the background processes ($\sigma_{\rm bkg}$)~\cite{preprint:Belle-II-Physics-Book}, the total background levels ($R_{\rm bkg}$) are calculated as
\begin{equation}
R_{\rm bkg} = \frac{\sum\limits_{\rm bkg}\sigma_{\rm bkg}\epsilon_{\rm bkg}}{(\sigma_{\rm ee}\epsilon_{\rm ee} + \sigma_{\rm \gamma\gamma}\epsilon_{\rm \gamma\gamma})}.
\label{equation: background level}
\end{equation}
\noindent The results are 0.07\% and 0.28\% in the Bhabha-dominant and digamma-dominant measurements, respectively.
Detailed background analysis shows that the background mainly arises from $u\bar{u}$, \tautau, and $d\bar{d}$ events in both measurements.

Inserting the values of $N_{\rm data}^{\rm obs}$, $\epsilon_{\rm ee}$, $\epsilon_{\rm \gamma\gamma}$, $\sigma_{\rm ee}$, $\sigma_{\rm \gamma\gamma}$, and $R_{\rm bkg}$ into the formula
\begin{equation}
L = \frac{N_{\rm data}^{\rm obs}}{(\sigma_{\rm ee}\epsilon_{\rm ee} + \sigma_{\rm \gamma\gamma}\epsilon_{\rm \gamma\gamma})(1+R_{\rm bkg})},
\label{equation: the complete formula for the measurement of integrated luminosities}
\end{equation}
the integrated luminosities are determined to be ($496.7 \pm 0.3$)~pb$^{-1}$ and ($493.1 \pm 0.7$)~pb$^{-1}$ in the Bhabha-dominant and digamma-dominant measurements, respectively.
Here, the uncertainties are statistical only.
In the two formulae above, the efficiencies $\epsilon_{\rm ee}$ and $\epsilon_{\rm \gamma\gamma}$ implicitly include an energy-sum-based ECL trigger efficiency of 100\% with a negligible uncertainty of $\mathcal{O}(0.01\%)$. This is evaluated using a radiative Bhabha data sample as the ratio of the events triggered by both ECL and CDC to all those triggered by CDC.

\section{Systematic uncertainties}
\label{section: Systematic uncertainties}
\begin{center}
\tabcaption{\label{table: Systematic uncertainties of the measured integrated luminosities.} Systematic uncertainties of the measured integrated luminosities. The second, third, and fourth columns list the uncertainties from the Bhabha-dominant, digamma-dominant, and combined measurements, respectively.}
\footnotesize
\begin{tabular}{cccc}
\toprule
Source & ee (\%) & $\gamma\gamma$ (\%) & ee + $\gamma\gamma$ (\%) \\
\hline
Cross section & $\pm$0.1 & $\pm$0.1 & $\pm$0.1 \\
CM energy &  $\pm$0.2 & $\pm$0.2 &  $\pm$0.2 \\
${\rm \theta}_{\rm cm}$ range & $\pm$0.0 &  $\pm$0.4 &  $\pm$0.1 \\
IP position & $\pm$0.2 & $\pm$0.1 &  $\pm$0.1 \\
ECL location & $\pm$0.2 & $\pm$0.2 &  $\pm$0.2 \\
MC statistics &  $\pm$0.1 & $\pm$0.1 &  $\pm$0.1 \\
Beam backgrounds &  $\pm$0.1 & $\pm$0.1 &  $\pm$0.1 \\
Cluster reconstruction &  $\pm$0.2 & $\pm$0.2 &  $\pm$0.2 \\
${E}_{\rm cm}$ distributions &  $\pm$0.1 & $\pm$0.1 &  $\pm$0.1 \\
${\rm \theta}_{\rm lab}$ distributions & $\pm$0.1 & $\pm$0.2 &  $\pm$0.1 \\
${\rm \theta}_{\rm cm}$ distributions & $\pm$0.3  & $\pm$0.3 &  $\pm$0.3 \\
${\rm \phi}_{\rm cm}$ distributions & $\pm$0.1  & $\pm$0.3 &  -- \\
Material effects & $-0.1$ & $+0.7$ &  +0.1 \\
Overlapping clusters & $\pm$0.1  & $\pm$0.1 &  $\pm$0.1 \\
Colliding backgrounds & $\pm$0.1 & $\pm$0.3 &  $\pm$0.1 \\
\hline
Quadrature sum & $\pm$0.6 & ${}_{-0.8}^{+1.1}$ &  $\pm$0.6 \\
\bottomrule
\end{tabular}
\end{center}

Table~\ref{table: Systematic uncertainties of the measured integrated luminosities.} summarizes the sources and values of the systematic uncertainties of the integrated luminosities measured above.
The systematic uncertainties are evaluated as follows.

The theoretical cross sections of Bhabha and digamma processes are evaluated with the {\sc BabaYaga@NLO} generator with a precision of 0.1\%~\cite{paper:BabaYaga@NLO-3,paper:BabaYaga@NLO-4}, which is taken as the relative systematic uncertainty in each measurement.

The CM energy is an essential input to the {\sc BabaYaga@NLO} generator for the evaluation of the signal cross sections and the generation of the signal events.
To check the impact of its uncertainty on the measured integrated luminosities, the two measurements are repeated with the CM energy increased/decreased by 0.1\%, which is roughly half the width of the \FourS resonance ($20.5\pm2.5$)~MeV~\cite{PDG2016} and is a conservative value for the energy uncertainty according to an analysis of the yield of $B$ mesons.
For each measurement, the larger of the changes in the integrated luminosity is taken as the associated uncertainty. The results are about 0.2\% for both measurements.
Additionally, since the rates of Bhabha and digamma processes vary comparatively slowly with energy, the impact of the uncertainty of the CM energy spread on the measured integrated luminosities is negligible.

The polar angle range of electrons and positrons for Bhabha events or photons for digamma events in the CM frame is another important input to the {\sc BabaYaga@NLO} generator.
The nominal signal MC samples are generated in the $\theta_{\rm cm}$ range 35$^{\circ}$--145$^{\circ}$. To check the impact of different $\theta_{\rm cm}$ ranges on the measured integrated luminosities, the two measurements are repeated with Bhabha and digamma events generated in the wider $\theta_{\rm cm}$ range 5$^{\circ}$--175$^{\circ}$.
For the Bhabha-dominant measurements, the results are consistent within the statistical uncertainties.
For the digamma-dominant measurements, the result changes by about 0.4\%, which is taken as the relative systematic uncertainty.

The actual position of the IP may deviate from the nominal position (0, 0, 0) as assumed in the MC simulation.
In a preliminary study with charged tracks, the average position and the width of the IP distribution over the whole data sample are determined to be ($-$0.4, 0.4, 0.3)~mm.
To investigate the impact of the deviation on the measured integrated luminosities, we repeat our measurements using a shifted position of the IP in the MC simulation. The shift used is ($-$0.4, $+$0.4, $+$0.3)~mm from the nominal position.
For the Bhabha-dominant and digamma-dominant measurements, the results change by about 0.2\% and 0.1\%, respectively.
In addition, the IP spread is calculated to be about (14~$\mu$m, 0.56~$\mu$m, 0.35~mm) with the optics parameters set for the $x$ and $z$ dimensions and observed for the $y$ dimension during Phase~2.
We perform a study with the IP spread, finding the IP spread only has a negligible impact on the measured integrated luminosities because its $x$ and $y$ components are small and its symmetry around the average position makes the effects in positive and negative directions essentially cancel.

The location of the ECL detector has an uncertainty of 0.5~mm in the $z$ direction.
In effect, this  uncertainty is equivalent to an uncertainty in the position of the IP, though they are from different sources.
To examine the impact of the uncertainty on the measured integrated luminosities, the two measurements are each repeated with two new sets of signal MC samples: one produced with the position of the IP changed from (0, 0, 0) to (0, 0, $+$0.5)~mm, another produced with the position of the IP changed to (0, 0, $-$0.5)~mm.
For both measurements, the larger change of the integrated luminosity is about 0.2\%, which is taken as the associated relative systematic uncertainty.
Besides the uncertainty in the $z$ direction, there is an uncertainty due to the rotation of the ECL sub-detector relative to the coordinate system. However, MC studies show that the impact of a rotation of 1~mrad in $\theta_{\rm lab}$ on the measured luminosities is negligible.

The relative systematic uncertainties due to the limited sizes of the signal MC samples are evaluated to be about 0.1\% for both measurements.

To examine the impact of beam background overlay on the measured integrated luminosities, MC samples without beam background overlay are produced and used to perform the two measurements.
The differences between the results obtained with and without the background overlay are taken as the systematic uncertainties.
The uncertainties are about 0.1\% for both measurements.
In addition, both MC samples with and without the background overlay demonstrate very good agreement with the data sample in the distributions of the number of ECL clusters after event selection. This indicates that beam backgrounds have only a negligible impact on the signal candidates, which have very clear signatures: two high energy clusters in the ECL barrel region and the back-to-back feature in the $\theta_{\rm cm}$ and $\phi_{\rm cm}$ projections.

We estimate the uncertainty due to ECL cluster reconstruction efficiencies using radiative Bhabha events. We find that the average relative difference between data and MC simulation in the efficiencies for the clusters in our selected events is about 0.1\%.
Since we have two clusters in both measurements, we take 0.2\% as the associated uncertainty.

The systematic uncertainties related to the distribution shapes of the energies, polar angles, and azimuthal angles of the ECL clusters are estimated by replacing the nominal requirements with alternatively more and less restrictive requirements.
For each distribution shape, the larger of the changes in integrated luminosity is taken as the associated uncertainty.
The requirements on the energies, polar angles, and acollinearity in polar angle in both the measurements are changed to
\begin{itemize}
\item (1.5) 2.5~GeV $<{\rm E}_{\rm cm}^{\rm max2}<{\rm E}_{\rm cm}^{\rm max1}<$ 5.62 (6.02)~GeV,
\item ($35.0^{\circ}$) $39.4^{\circ}$ $<\theta_{\rm lab}^{{\rm max}1}$, $\theta_{\rm lab}^{{\rm max}2}<$ $118.4^{\circ}$ ($124.6^{\circ}$), and
\item $|\theta_{\rm cm}^{\rm max1}+\theta_{\rm cm}^{\rm max2}-180^\circ|<$ $2.5^{\circ}$ ($7.5^{\circ}$);
\end{itemize}
the requirement on the acollinearity in azimuthal angle in the Bhabha-dominant measurement is changed to
\begin{itemize}
\item ($1.5^{\circ}$) $3.5^{\circ}$ $<||\phi_{\rm cm}^{\rm max1}-\phi_{\rm cm}^{\rm max2}|-180^\circ|<$ $12^{\circ}$ ($14^{\circ}$);
\end{itemize}
and the requirement on the acollinearity in azimuthal angle in the digamma-dominant measurement is changed to
\begin{itemize}
\item $||\phi_{\rm cm}^{\rm max1}-\phi_{\rm cm}^{\rm max2}|-180^\circ|<$ $1.5^{\circ}$ ($3.5^{\circ}$).
\end{itemize}
Here, the values inside and outside the parentheses correspond to the looser and tighter alternative requirements, respectively.
The estimated systematic uncertainties obtained by changing requirements on these parameters are listed in Table~\ref{table: Systematic uncertainties of the measured integrated luminosities.}.

A photon, electron or positron may interact while traversing the material in the VXD region.
As mentioned in Section \ref{section: Monte Carlo simulation}, the material is not fully included in the simulation model, and hence the material effects differ between the data and MC samples.
To check the impact of the difference on the measured integrated luminosities, the two measurements are repeated  with a new set of Bhabha and digamma MC samples produced with the vertex detectors removed from the simulation and reconstruction programs.
Corresponding to the change of signal MC samples, the integrated luminosity obtained in the Bhabha-dominant measurement increases by about 0.42\%, while that obtained in the digamma-dominant measurement decreases by about 3.5\%.
As described in Section \ref{section: Monte Carlo simulation}, the unsimulated material is  estimated to be 20\% of the simulated material, and therefore we take $-$20\% instead of 100\% of the resulting changes as the associated systematic uncertainties.
The relative uncertainties are estimated to be $-0.1$\% and $+0.7$\% for the Bhabha-dominant and digamma-dominant measurements, respectively.
Here, the uncertainties are signed and show the reduction in the difference between the Bhabha-dominant and digamma-dominant measurements.

A photon, electron or positron may also interact with material while traversing the CDC outer wall and the TOP detector, resulting in two nearby ECL clusters.
Because we preferentially select events that do not contain nearby clusters, imperfect modeling of this process could lead to a systematic uncertainty.
We evaluate the uncertainty by repeating the two measurements with the selection criteria supplemented by requirements dedicated to select events with pairs of nearby clusters.
With the extra requirements applied, the change of the result is less than 0.1\% for both measurements, which is conservatively taken as the relative systematic uncertainty.

Besides the signal events, a small fraction of background events survive the event selection.
We take 100\% of the total background levels as the associated systematic uncertainties, which are about 0.1\% and 0.3\% in the Bhabha-dominant and digamma-dominant measurements, respectively.

Assuming that the individual uncertainties are independent and adding them in quadrature yields total relative systematic uncertainties of 0.6\% and ${}_{-0.8}^{+1.1}$\% for the Bhabha-dominant and digamma-dominant measurements, respectively.
Including these total systematic uncertainties, the integrated luminosities are ($496.7 \pm 0.3 \pm 3.0$)~pb$^{-1}$ and ($493.1 \pm 0.7 _{-4.0}^{+5.4}$)~pb$^{-1}$ for the Bhabha-dominant and digamma-dominant measurements, respectively.
The systematic uncertainties dominate in both measurements.
Accounting for the correlations between the uncertainties for the Bhabha- and digamma-dominant measurements, the ratio of the two luminosities is determined to be $1.007\pm0.002\pm0.008$, indicating agreement between the two results.

As can be seen from Section \ref{section: Event selection}, the signal candidates in the Bhabha-dominant and digamma-dominant measurements are separated by the border $||\phi_{\rm cm}^{\rm max1}-\phi_{\rm cm}^{\rm max2}|-180^\circ|=2.5^{\circ}$.
To get the combined result of the two measurements, we repeat a measurement with the merged requirement $||\phi_{\rm cm}^{\rm max1}-\phi_{\rm cm}^{\rm max2}|-180^\circ|<13^{\circ}$.
In this measurement, systematic uncertainties are estimated with the same methods used in the two separate measurements, and the results are listed in the fourth column of Table \ref{table: Systematic uncertainties of the measured integrated luminosities.}.

Because most of the uncertainty sources are the same for the two separate measurements and Bhabha events dominate the signal candidates in the combined measurement ($\frac{\sigma_{\rm ee}\epsilon_{\rm ee}}{\sigma_{\rm \gamma\gamma}\epsilon_{\rm \gamma\gamma}} \approx 6.7$), almost all of the systematic uncertainties are equal to their counterparts in the Bhabha-dominant measurement at the order of 0.1\%.
The uncertainty associated with ${\rm \phi}_{\rm cm}$ distributions is negligible, since $||\phi_{\rm cm}^{\rm max1}-\phi_{\rm cm}^{\rm max2}|-180^\circ|<13^{\circ}$ is a relatively loose requirement.
The uncertainty related to material effects is estimated to be +0.1\%, mainly because of the cancellation of the corresponding uncertainties in the two separate measurements with the associated numbers of signal candidates as weights. With the systematic uncertainties, the combined result is calculated to be ($496.3 \pm 0.3 \pm 3.0$)~pb$^{-1}$, which is nearly the same as in the Bhabha-dominant measurement.
We take the combined result as the final result in this work.

\section{Conclusions}
\label{Conclusions}
The integrated luminosity of the first data sample collected with the Belle~II detector at SuperKEKB during Phase~2 is measured using ECL information with Bhabha and digamma events.
The result obtained in the Bhabha-dominant measurement is consistent with that obtained in the digamma-dominant measurement.
Combining the two measurements, we determine the integrated luminosity to be ($496.3 \pm 0.3 \pm 3.0$)~pb$^{-1}$, where the first uncertainty is statistical and the second is systematic.

The result will be used in the early studies with the Phase~2 data at Belle~II, particularly in the searches for new physics in the dark sector, in which Belle~II expects to achieve good sensitivities owing to the dedicated triggers for single photon and low multiplicity events~\cite{paper:dark_sector}.
Using ECL information alone, this work builds a foundation for future luminosity measurements in the Belle~II experiment, in which we will incorporate the information obtained by other sub-detectors, particularly the CDC, to select signal events.

\ \\

\acknowledgments
{
We thank the SuperKEKB group for the excellent operation of the
accelerator; the KEK cryogenics group for the efficient
operation of the solenoid; and the KEK computer group for
on-site computing support.
This work was supported by the following funding sources:
Science Committee of the Republic of Armenia Grant No. 18T-1C180;
Australian Research Council and research grant Nos.
DP180102629,
DP170102389,
DP170102204,
DP150103061,
FT130100303,
and
FT130100018;
Austrian Federal Ministry of Education, Science and Research, and
Austrian Science Fund No. P 31361-N36;
Natural Sciences and Engineering Research Council of Canada, Compute Canada and CANARIE;
Chinese Academy of Sciences and research grant No. QYZDJ-SSW-SLH011,
National Natural Science Foundation of China and research grant Nos.
11521505,
11575017,
11675166,
11761141009,
11705209,
and
11975076,
LiaoNing Revitalization Talents Program under contract No. XLYC1807135,
Shanghai Municipal Science and Technology Committee under contract No. 19ZR1403000,
Shanghai Pujiang Program under Grant No. 18PJ1401000,
and the CAS Center for Excellence in Particle Physics (CCEPP);
the Ministry of Education, Youth and Sports of the Czech Republic under Contract No.~LTT17020 and
Charles University grants SVV 260448 and GAUK 404316;
European Research Council, 7th Framework PIEF-GA-2013-622527,
Horizon 2020 Marie Sklodowska-Curie grant agreement No. 700525 `NIOBE,'
Horizon 2020 Marie Sklodowska-Curie RISE project JENNIFER grant agreement No. 644294,
Horizon 2020 ERC-Advanced Grant No. 267104, and
NewAve No. 638528 (European grants);
L'Institut National de Physique Nucl\'{e}aire et de Physique des Particules (IN2P3) du CNRS (France);
BMBF, DFG, HGF, MPG and AvH Foundation (Germany);
Department of Atomic Energy and Department of Science and Technology (India);
Israel Science Foundation grant No. 2476/17
and
United States-Israel Binational Science Foundation grant No. 2016113;
Istituto Nazionale di Fisica Nucleare and the research grants BELLE2;
Japan Society for the Promotion of Science,  Grant-in-Aid for Scientific Research grant Nos.
16H03968,
16H03993,
16H06492,
16K05323,
17H01133,
17H05405,
18K03621,
18H03710,
18H05226,
19H00682, 
26220706,
and
26400255,
the National Institute of Informatics, and Science Information NETwork 5 (SINET5),
and
the Ministry of Education, Culture, Sports, Science, and Technology (MEXT) of Japan;
National Research Foundation (NRF) of Korea Grant Nos.
2016R1\-D1A1B\-01010135,
2016R1\-D1A1B\-02012900,
2018R1\-A2B\-3003643,
2018R1\-A6A1A\-06024970,
2018R1\-D1A1B\-07047294,
2019K1\-A3A7A\-09033840,
and
2019R1\-I1A3A\-01058933,
Radiation Science Research Institute,
Foreign Large-size Research Facility Application Supporting project,
the Global Science Experimental Data Hub Center of the Korea Institute of Science and Technology Information
and
KREONET/GLORIAD;
Universiti Malaya RU grant, Akademi Sains Malaysia and Ministry of Education Malaysia;
Frontiers of Science Program contracts
FOINS-296,
CB-221329,
CB-236394,
CB-254409,
and
CB-180023, and the Thematic Networks program (Mexico);
the Polish Ministry of Science and Higher Education and the National Science Center;
the Ministry of Science and Higher Education of the Russian Federation,
Agreement 14.W03.31.0026;
Slovenian Research Agency and research grant Nos.
J1-9124
and
P1-0135;
Agencia Estatal de Investigacion, Spain grant Nos.
FPA2014-55613-P
and
FPA2017-84445-P,
and
CIDEGENT/2018/020 of Generalitat Valenciana;
Ministry of Science and Technology and research grant Nos.
MOST106-2112-M-002-005-MY3
and
MOST107-2119-M-002-035-MY3,
and the Ministry of Education (Taiwan);
Thailand Center of Excellence in Physics;
TUBITAK ULAKBIM (Turkey);
Ministry of Education and Science of Ukraine;
the US National Science Foundation and research grant Nos.
PHY-1807007 
and
PHY-1913789, 
and the US Department of Energy and research grant Nos.
DE-AC06-76RLO1830, 
DE-SC0007983, 
DE-SC0009824, 
DE-SC0009973, 
DE-SC0010073, 
DE-SC0010118, 
DE-SC0010504, 
DE-SC0011784, 
DE-SC0012704; 
and
the National Foundation for Science and Technology Development (NAFOSTED)
of Vietnam under grant No 103.99-2018.45.
}

\end{multicols}

\vspace{-1mm}
\centerline{\rule{80mm}{0.1pt}}
\vspace{2mm}

\begin{multicols}{2}

\end{multicols}

\clearpage
\end{document}